\RequirePackage{fix-cm}
\documentclass{svjour3}                     
\smartqed  
\usepackage{graphicx}
\usepackage{mathptmx}   
\usepackage{colortbl}
\usepackage{amsmath,amsfonts}
\usepackage{algorithmic}
\usepackage{graphicx}
\usepackage{textcomp}
\usepackage{float}
\usepackage{subcaption} 
\usepackage{multirow}
\usepackage[inline]{enumitem}
\usepackage{pbox}
\usepackage{booktabs}

\usepackage{lineno,hyperref}
\usepackage{soul}
\usepackage{color}
\usepackage{xcolor,framed,graphicx}
\definecolor{beaublue}{rgb}{0.74, 0.83, 0.9}
\usepackage{numprint}
\npthousandsep{,}
\npdecimalsign{.}
\usepackage[title]{appendix}
\journalname{Empirical Software Engineering}
\begin{document}

\title{Semantically-enhanced Topic Recommendation Systems for Software Projects}
\titlerunning{Semantically-enhanced Topic Recommendation Systems}

\author{Maliheh Izadi \and
        Mahtab Nejati \and
        Abbas Heydarnoori}
\authorrunning{Izadi et al.}

\institute{M. Izadi \at
            Technical University of Delft, Netherlands \\
            \email{m.izadi@tudelft.nl}
          \and
          M. Nejati \at
            University of Waterloo, Canada \\
            \email{mahtab.nejati@uwaterloo.ca}
          \and
          A. Heydarnoori \at
          Bowling Green State University, USA\\
          Dr. Heydarnoori is also a corresponding author for this work.\\
          \email{aheydar@bgsu.edu}
}

\date{Received: date / Accepted: date}
\maketitle

\begin{abstract}
Software-related platforms 
such as GitHub and Stack Overflow, 
have enabled their users to collaboratively 
label software entities with a form of metadata called topics. 
Tagging software repositories with relevant topics
can be exploited for facilitating various downstream tasks. 
For instance, a correct and complete set of topics 
assigned to a repository can increase its visibility.
Consequently, this improves the outcome of tasks 
such as browsing, searching, navigation, and organization of repositories.
Unfortunately, assigned topics are usually highly noisy, 
and some repositories do not have well-assigned topics.
Thus, there have been efforts on recommending topics for software projects, 
however, the semantic relationships among these topics have not been exploited so far.

In this work, we propose two recommender models for tagging software projects
that incorporate the semantic relationship among topics.
Our approach has two main phases;
(1) we first take a collaborative approach to curate a dataset of quality topics 
specifically for the domain of software engineering and development.
We also enrich this data with the semantic relationships among these topics
and encapsulate them in a knowledge graph we call \emph{SED-KGraph}.
Then, (2) we build two recommender systems; 
The first one operates only based on 
the list of original topics assigned to a repository 
and the relationships specified in our knowledge graph.
The second predictive model, however, 
assumes there are no topics available for a repository, 
hence it proceeds to predict the relevant topics 
based on both textual information of a software project 
(such as its README file), and SED-KGraph.

We built SED-KGraph in a crowd-sourced project with $170$ contributors 
from both academia and industry. 
Through their contributions, we constructed SED-KGraph 
with \numprint{2234} carefully evaluated relationships 
among $863$ community-curated topics.
Regarding the recommenders' performance, 
the experiment results indicate that our solutions 
outperform baselines that neglect the semantic relationships among topics
by at least $25\%$ and $23\%$ in terms of 
Average Success Rate and Mean Average Precision metrics, respectively.
We share SED-KGraph, as a rich form of knowledge 
for the community to re-use and build upon.
We also release the source code of 
our two recommender models, KGRec and KGRec+.~\footnote{\url{https://github.com/mahtab-nejati/KGRec}}

\end{abstract}

\section{Introduction}\label{intro}
Software engineers and developers explore Software Information Sites
such as GitHub and Stack Overflow
to find interesting software components 
tailored to their needs,
to reuse source code, 
to find answers to their programming questions and many more.
However, the sheer number of software entities (projects, questions, etc.) 
hosted on these sites 
hinders efficient searching, retrieving, navigating, and categorizing said entities. 
For instance, GitHub currently hosts 
more than $240$ million software projects.~\footnote{January 2022, \url{https://github.com/search}}
Many of these projects share common characteristics 
such as similar objectives and functionalities. 
With the continuous growth of these platforms,
more advanced automatic solutions are needed
to improve the retrieval of relevant software projects.
Existing techniques for better organization, 
documentation or and retrieval of software entities include 
various types of recommender systems~\cite{xia2013tag,xia2015tagcombint,vargas2015automated,zhou2017scalable,wang2018entagrec++,liu2018fasttagrec,izadi2021topic,izadi2022predicting,mazrae2021automated},
similar repository retrieval~\cite{mcmillan2012detecting,thung2012detecting,zhang2017detecting}, 
and project clustering~\cite{escobar2015unsupervised,zhang2019higitclass,reyes2016deep,yang2016cut}.
Topics, also known as tags,
are a form of concise yet highly valuable metadata, 
that enrich software entities with human knowledge.
Topics annotate an entity based on its core concepts. 
A software topic encompasses the key features of a repository
including which category it belongs to, 
its main programming language, 
its intended audience, 
its user interface, and more. 
Topics complement textual descriptions of repositories
as they highlight their main aspects with explicit and short tokens. 
Thus, topics can greatly help 
with the visibility of relevant entities to user queries. 
Consequently, they are used for improving 
the organization and retrieval of software repositories.

Topics can convey information in two ways;
\emph{explicitly} as stand-alone sources of information,
and \emph{implicitly} through their semantic connections to one other.
The former has been extensively exploited 
to build topic recommendation systems~\cite{xia2013tag,di2020multinomial,dirocco2020topfilter,izadi2021topic}.
For the latter, consider the topic \texttt{angular} 
which refers to an open-source web application framework.~\footnote{\url{https://angular.io/}}
As Angular provides functionality for front-end development,
a programmer can almost immediately relate this topic 
to the \texttt{frontend} or \texttt{web-development} topic.
Thus, there exist implicit links 
between topics \texttt{angular}, \texttt{frontend}, and \texttt{web-development}.
In practice, repository owners 
-probably due to a lack of motivation-  
neglect tagging their projects with sufficient topics.
Implicit connections mentioned above, 
can be utilized to track missing information,
complement such incomplete topic sets, 
and consequently, improve the visibility of a given repository.
They can also help recommendation systems suggest more accurate topic lists. 
However, the semantic relationships among software engineering topics 
and their impact on the performance of such predictive models 
are not properly explored yet.
In this study,
we strive to build more advanced recommendation systems 
for predicting key topics of software repositories
through exploiting these relationships.

As much as topics and the relationships among them appear enticing 
for improving automated information retrieval-based tasks, 
there are several challenges for employing them in real-world scenarios, 
including the well-known \emph{tag explosion} phenomenon~\cite{golder2006usage} 
and the problem we call \emph{tangled topics}.
As users are free to define topics in the free-format text,
they can create differently-written yet synonymous topics for any given concept, 
as well as compound and personal topics.
This freedom results in an explosion of tags.
That is when the set of topics exceedingly grows in number 
to the point that the sheer multitude of topics defeats their intended purposes. 
Furthermore, inspecting topics assigned by users,
we came upon many tangled topics.
These are compound topics that bundle multiple atomic concepts into a single tag 
and treat this tag as a distinct concept.
Note that compound topics 
which communicate an atomic concept do exist, 
e.g., \texttt{single-page-application} is an atomic topic
which should not be further dissected into multiple topics.
However, a compound topic such as \texttt{java-library} 
can easily be broken down into its constituent atomic concepts 
without losing any semantic content.
The same can happen by adding adjectives to the existing atomic topics, such as \texttt{small-library} or \texttt{big-library} and the situation exacerbates when \texttt{small-java-library} is also considered a unique concept.
Unfortunately, some models redundantly recommend 
such topics together for a given repository.
We believe that learning and recommending tangled topics 
adds little value 
when an entity is already assigned 
their atomic constituent topics 
while increasing the size and complexity of the topic set.
As a result,
to build an enhanced recommender system,
we need to address both tag explosion and tangled topics problems 
through carefully assessing the input set of topics.
In an attempt to resolve the above challenges
and to collect a set of quality topics, 
GitHub has commenced a crowd-sourced project to feature a set of 
community-curated Software Engineering and Development (SED) topics.~\footnote{\url{https://github.com/github/explore}}
At the time of commencing this study, 
this project curated $389$ topics over the course of almost three years.
This set of GitHub's featured topics contains valuable explicit information,
however, semantic relationships among topics are missing.
In this study, we take this seed as an initial set for topic collection
and build upon it by acquiring more high-quality topics, 
and annotating them with semantic information based on human knowledge.

The next challenge is to properly store the high-quality SED topics 
along with their semantic connections.
Knowledge graphs (KG) are a viable solution to this problem.
More specifically, such relationships can be modeled in the form of relation triples 
$\langle$ \textit{subject, verb-phrase, object} $\rangle$.
Take the previous example, 
we can store two types of relations 
as $\langle$\texttt{angular}, is-a, \texttt{framework}$\rangle$,
and $\langle$\texttt{angular}, provides-functionality, \texttt{frontend}$\rangle$.
KGs have been shown useful in different tasks 
such as information retrieval, recommendation, question answering, and search results ranking~\cite{zou2020survey}.
As a prominent example, Google's KG is used to enhance its search engine results.
They have also been widely used in domain-specific applications 
for medical, financial, news, social networks, etc. purposes~\cite{zou2020survey}
as well as software engineering~\cite{li2018api,chen2019bug}.
A KG of SED  topics 
can improve the performance of topic-dependent tasks 
based on the topics assigned to the entities.
In addition, such a KG can also be utilized 
as a structured knowledge base for the community to query, navigate, 
and perform an exploratory search.
Hence, we aim to store the semantic information along with our topics in a KG, 
which we then feed into our predictive model to recommend better topics.

A domain-specific KG 
can be built automatically through processing domain knowledge,
manually with the help of domain experts, or in a hybrid manner.
In the software engineering domain,
there exist a few semi- or fully-automatic approaches to build KGs~\cite{zhao2017hdskg,li2018api,chen2019bug,sun2019knowhow,sun2020taskoriented}. 
Zhao et al. propose \textit{HDSKG} using a  
semi-automatic approach to construct 
a KG of SED topics~\cite{zhao2017hdskg}.
Although HDSKG aims to minimize the manual effort 
that goes into the construction of a KG, 
it obsessively chunks noun phrases. 
This leads to the introduction of numerous tangled topics in this KG.
Moreover, the knowledge scope acquired using fully-automatic approaches 
tends to be restricted to specific aspects/technologies
such that the concepts can be predetermined or easily extracted.
Construction of a KG of SED topics at the scope of this study 
is much more challenging
due to the diversity of topics, their types of relationships, 
and the different abstraction levels of topics.
Not to mention the data sparsity on particular topics, 
and data scatteredness across the web and multiple sources
which cause duplicate, incomplete, and incorrect data~\cite{fathalla2018eventskg}.
Consequently, we take a mostly manual approach 
in conjunction with automation techniques for facilitating knowledge acquisition and evaluation.

In the first phase of our approach,
using the contributions of $170$ SE experts 
from both academia and industry,
we acquire high-quality SED topics,
extract their semantic relationships,
evaluate this information,
and store them in a domain-specific KG 
we call \textit{SED-KGraph}.
We developed an online platform on which 
we partially automated the growth of SED-KGraph 
using the help of our contributors in multiple snapshots.
We expand the set of GitHub's featured topics 
to a more comprehensive and inclusive one.
To guarantee the consistency of SED-KGraph, 
we centrally coordinate the expansion of this KG.
By capturing the semantic relationships among topics in a KG, 
utilizing SED-KGraph can potentially improve the performance of solutions to numerous software community problems 
such as software entity classification, automated labeling, navigation, search, etc.
While our approach to KG construction is a manual one, 
similar to \texttt{FreeBase}~\cite{bollacker2008freebase} 
which has been utilized in automated tasks in other studies~\cite{dong2015question,xu2016question,yao2014information}, 
SED-KGraph too can pave the way for numerous automated topic-dependent tasks, 
while it continues to grow further over time.

In the second phase, 
we propose recommendation systems for two scenarios; 
(1) \textit{KGRec}, a topic recommender system to predict \textit{missing topics}, 
the topics relevant to the entities but not assigned to them by users.
Correctly predicting the missing topics 
improves the completeness of the set of topics assigned to each project,
which has been shown as an important factor 
in performance of solutions to topic-dependent tasks~\cite{held2012learning}.
We build KGRec purely based on SED-KGraph 
through the application of spreading activation techniques.
(2) Next, we build upon KGRec 
by adding a Machine Learning-based (ML) component 
to the model and proposing \textit{KGRec+},
a fully automated topic prediction model.
KGRec+ works based on both the projects' textual data 
and the knowledge captured in SED-KGraph.

We demonstrate that the recommender systems based on KGRec 
outperform the ones based on \textit{TopFilter}, 
the state-of-the-art technique for relevant topic prediction~\cite{dirocco2020topfilter}, 
especially when the set of initial topics assigned to the project is limited in number.
Our contributions are as follows:
\begin{itemize}
    \item We develop and evaluate two topic recommenders 
    that outperform the competing approaches by $23\%$ to $151\%$
    in terms of Mean Average Precision (MAP) score.
    \item We collaboratively augment the set of GitHub featured topics
    with $393$ community-suggested topics.
    Furthermore, we present SED-KGraph to capture the semantic relationships among atomic 
    and semantically unique SED topics to improve the performance of topic recommenders.
    We engage $170$ practitioners and researchers from $16$ technology-based companies 
    and $11$ universities in the expansion and validation of SED-KGraph.
    The resultant KG consists of $863$ topics, 
    \numprint{2234} verified relationships, and $13$ relation types.
    \item We publicly share our two main software artifacts; 
    the data component (SED-KGraph), 
    and the model component (KGRec, KGRec+) along with their source code 
    for use by the SE community.~\footnote{\url{https://github.com/mahtab-nejati/KGRec}}
\end{itemize}

In the following, 
we first define the problem formally.
Next, we present the approach in \autoref{sec:approach}, 
experiments' settings in \autoref{sec:settings}, 
and our results in \autoref{sec:resutls}.
We then discuss the results, lessons learned, 
and possible applications and implications of this work.
Finally, we discuss the threats to the validity of this study 
and review the related work around the study.

\section{Problem Definition}\label{sec:problem}
GitHub hosts millions of repositories
$S = \{r_1, r_2, .., r_n\}$, 
where $r_i$ denotes a single software project.
Each repository may contain various types of information 
such as a description, README files, wiki pages, and source code files. 
Each project may include a set of topics $T = \{t_1, t_2, ...,t_m\}$, 
where $m$ is the number of assigned topics to a repository. 
Our goal is to 
(1) augment the initial set of topics assigned to a given repository $r_i$, 
or (2) recommend a set of topics from scratch for a given topic-less repository $r_j$.
In both cases, we aim to enhance recommender models 
using semantic relationships among high-quality topics.

\section{Approach}\label{sec:approach}
\begin{figure}[bt]
    \centering
    \includegraphics[width=\linewidth]{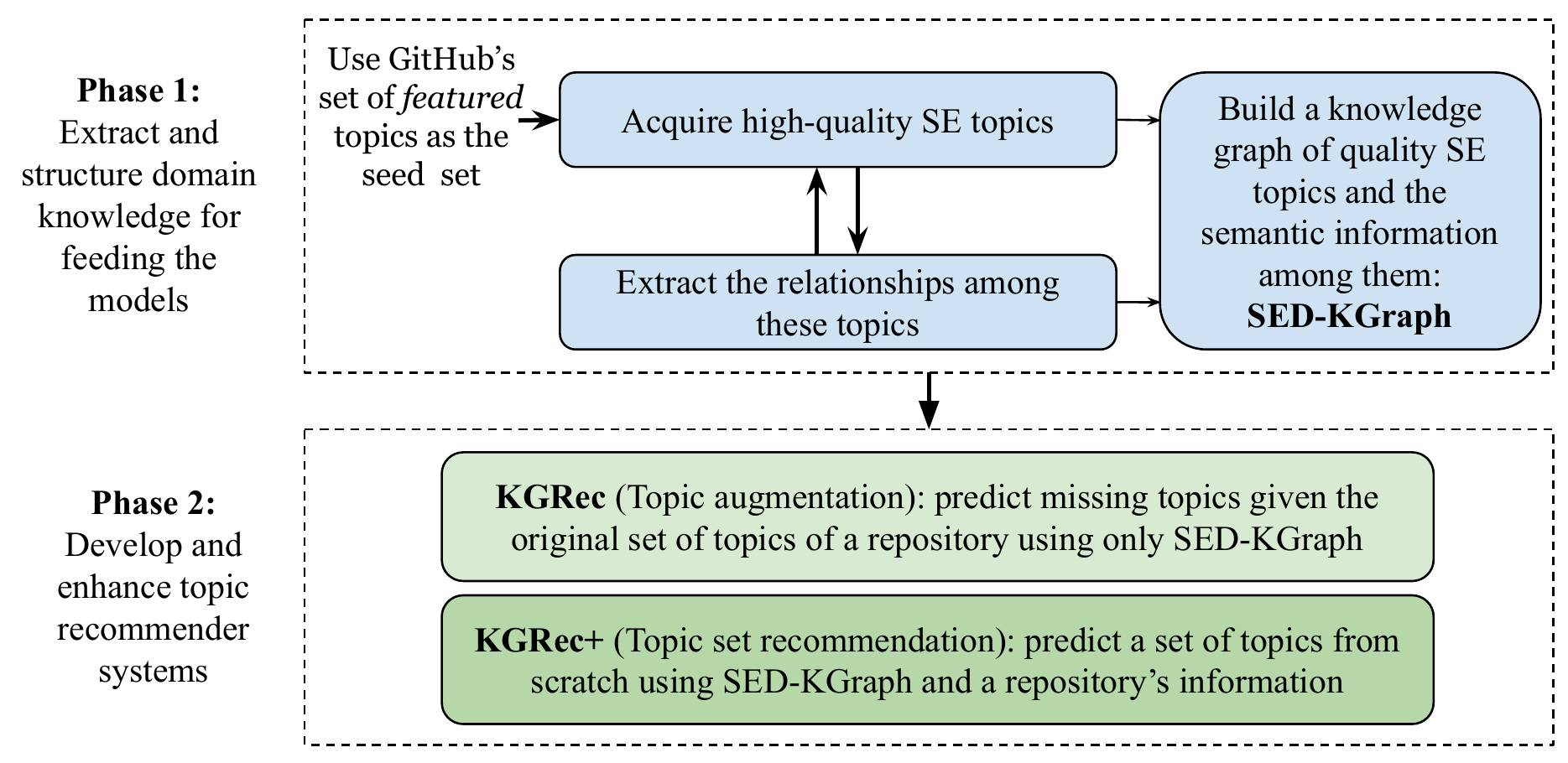}
    \caption{Overall workflow of our proposed approach}
    \label{fig:overall_approach}
\end{figure}

Our approach consists of two main phases; 
(1) acquire and store high-quality topics and the semantic relationships among them, 
(2) build stronger recommenders exploiting the semantic source of information.
In the first phase, 
we exploit explicit human knowledge in the domain 
to procure rich input data for our topic prediction models.
In the next phase, we propose two recommenders; 
\textit{KGRec} is a topic \textit{augmentation} model
which is used when an initial set of topics is already 
assigned to a given project which we aim to extend, 
i.e., predict the missing topics only based on the original set.
Finally, we stack this model on top of a ML-based component, 
building \textit{KGRec+}, an automated \textit{topic set} recommender system.
KGRec+ eliminates the need for the initial set of topics
and takes the textual data on the project and SED-KGraph as input.
\autoref{fig:overall_approach} 
depicts the overall workflow of our approach.
In the following, we provide more details on the proposed approach.

\subsection{Phase 1: KG Construction}
\begin{figure}[bt]
    \centering
    \includegraphics[width=0.9\linewidth]{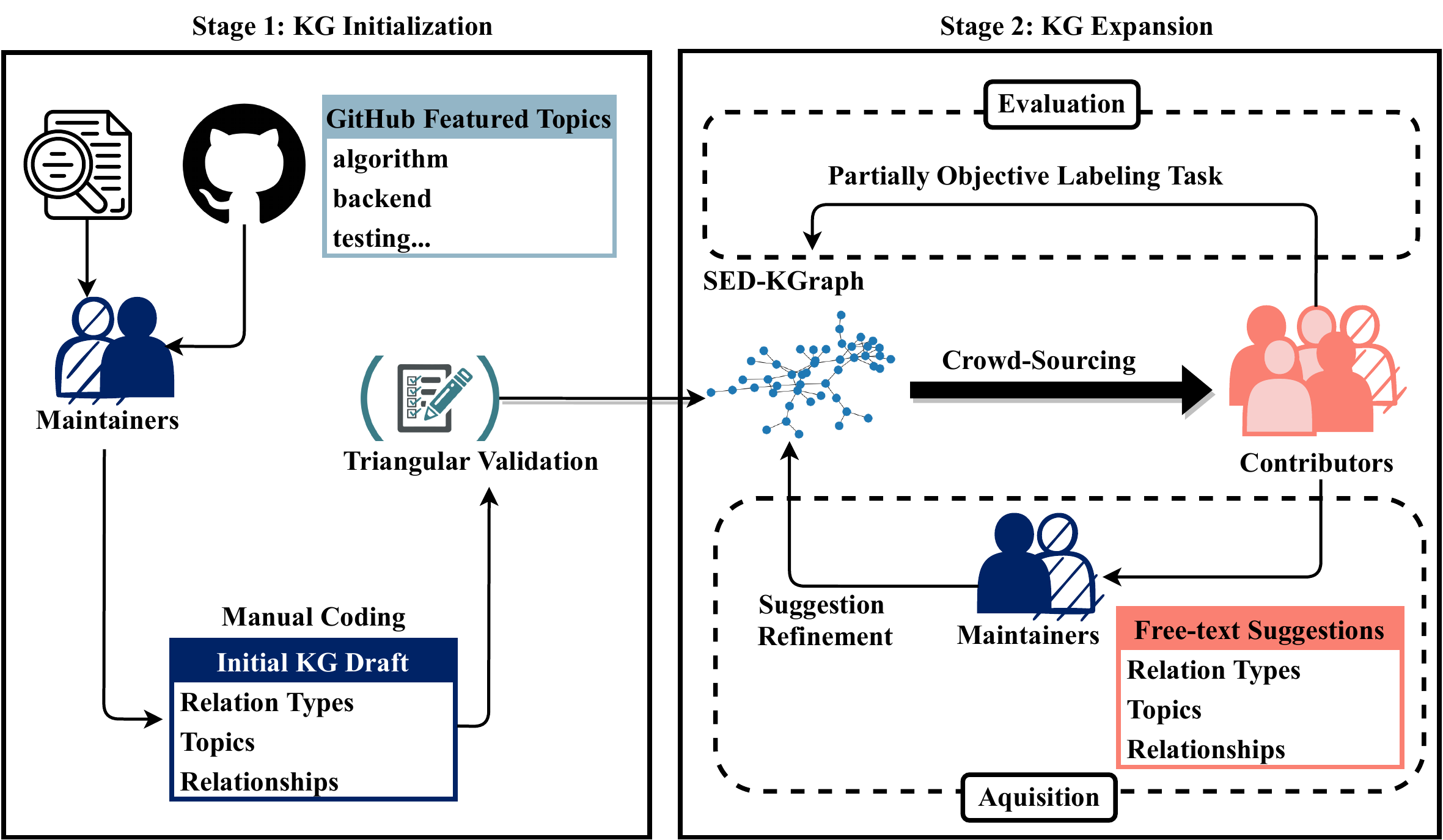}
    \caption{KG construction}
    \label{fig:kg_approach}
\end{figure}

As part of our approach for building better recommender systems, 
we need to utilize high quality software engineering topics and their semantic relationships.
KGs are a viable solution 
to store such information in a structured format.
In this section, we lay out the methodology through 
which we construct and evaluate such KG.
We utilize a crowd-sourcing technique 
to build the SED-KGraph in a two-step process 
with the second step being an on-going and continuous expansion phase.
To this end, we design an online platform for SED-KGraph's growth 
through community contributions.
Throughout the process, individuals are involved 
in one of the two roles of \textit{maintainer} or \textit{contributor}.
The first two authors take on the role of maintainers 
and the participants of the second step are the contributors.
\autoref{fig:kg_approach} demonstrates 
the overall process of KG construction.

\subsubsection{Initialization}\label{sec:kg_approach_initialization}
Maintainers initialize SED-KGraph 
with a set of topics, relation types, and relationships 
in a \textit{manual coding} process~\cite{wagner2015}.
They incorporate the \textit{triangular validation}~\cite{wagner2015} process 
in which coders cross-evaluate the coding results.
To initialize SED-KGraph, 
maintainers studied the $389$ topics featured provided by GitHub
and used it as the seed set.
For each topic, maintainers studied the information available on GitHub about it, 
as well as the top projects on GitHub (based on the number of stargazers) labeled with it.
Moreover, maintainers also searched each topic online to glean more knowledge on them.
They referenced these projects to make sure their understanding of the topic 
matched with the usage of the topic in the community.

After acquiring an overall insight into the topics, 
maintainers defined the relationships among them in a manual coding process.
They first discussed the possible types of relationships among the topics 
and decided on four basic, yet strongly effective ones as a primary set of relation types.
Note that this decision was made with the \textit{conciseness} feature in mind, 
knowing that the primary set of relation types is not comprehensive.
To minimize the impact of the maintainers' subjectiveness 
on the resultant KG, 
they take note to keep the initial structure minimal. 
by making the initial draft as concise as possible.
Yet, the maintainers look to include as many distinct atomic topics 
as possible to cover the diverse range of SED topics.
Then, they defined the relationships in an iterative manual coding process.
In this process, each maintainer iterated over the set of topics several times, 
each time defining and/or correcting the relationships.
Maintainers also incorporated triangular validation 
into the process to validate the relationships.
They reviewed the relationships defined by each other
to validate their correctness and effectiveness.
In case of a disagreement, 
maintainers discussed their reasons 
for approving/disapproving of a relationship 
and made the final decision on the relationship's correctness together.
If a consensus was not made, 
they included the relationships of conflict in the initial KG 
to allow the contributors 
to deliver the final verdict on the correctness of 
these relationships as a third jury.

\subsubsection{Expansion}\label{sec:kg_approach_expansion}
After establishing the first draft of SED-KGraph, 
the defined relationships needed to be evaluated.
Furthermore, acquiring the community's knowledge 
on the topics and the relationships among them 
would help expand SED-KGraph 
into a more comprehensive KG.
During the Expansion step, 
two separate tasks run simultaneously: 
\textit{a) Evaluation} of the previously defined relationships, 
and \textit{b) Acquisition} of new community knowledge on topics, 
relation types, and relationships among the topics.
To facilitate this step,
we deployed an online platform
through which contributors engage in evaluation and expansion of the KG.
Contributors review the relationships among topics 
and submit their suggestions in free-text forms.
In the online platform, 
contributors are allowed to skip unfamiliar topics 
and choose topics of their interest to contribute to.
This measure is taken to avoid mandating contributions
when contributors are not familiar with a concept.

Finally, for this KG to be valid at all times, 
a \textit{continuous expansion} method is required to maintain and update it with new knowledge.
This is because the SED fields are dynamic, 
i.e., new topics frequently emerge and/or evolve 
such that the previously defined relationships can be affected.
Moreover, we can not obviate the probability of unnoticed topics, 
i.e., topics that are insignificant or unknown to the contributors at the moment 
but grow into popularity over time.
Therefore, the platform later evolved for maintenance purposes.

\paragraph{Evaluation}\label{sec:kg_approach_expansion_evaluation}
In the KG Expansion step, 
contributors validate the correctness of 
the already defined relationships in SED-KGraph.
They evaluate all the relationships 
defined during the Initialization step and Acquisition task.
We achieve this through a 
\textit{partially objective labeling task}~\cite{alonso2014crowdsourcing}, 
a type of crowd-sourced labeling task 
in which the label of the subject to the task 
is determined based on inter-rater agreement.
That is, the subject $\langle relationship \rangle$ 
is assigned the label (``True" or ``False") 
which the majority of raters (contributors) have given the subject.
Contributors validate the correctness of each relationship 
by labeling it with ``True" 
and disapprove of the relationship 
by labeling it with ``False".
In the end, we consider the relationships as ``approved" 
and add them to SED-KGraph only if 
the majority of the contributors 
who reviewed the relationship have labeled it with ``True".
Otherwise, the relationship is disposed of.
Therefore, to determine the objective label for each relationship, 
we need votes from at least three contributors.

\paragraph{Acquisition}\label{sec:kg_approach_expansion_acquisition}
The expansion of SED-KGraph solely depends on community contributions.
To expand SED-KGraph such that it covers 
the currently missing SED topics and the relationships among them,
we ask the contributors to provide relationship suggestions 
for each of the topics, they review through free-text forms.
We also allow for new topic and relation type definitions
As contributors submit their suggestions in the free-format text, 
there is a chance for tag explosion and/or tangled topics.
Moreover, the semantic uniqueness 
of the relation types and topics is also at risk, 
leading to redundancy/duplication.
To mitigate such occurrences, we implement policies and functionalities, 
some of which are (described in Section~\ref{sec:platform}) 
into the online platform over which SED-KGraph is maintained.
These policies and functionalities are designed 
such that they give the contributors autonomy 
in expanding the KG while ensuring 
that the integrity and consistency of the KG are preserved.
Through initial snapshots,
maintainers inspected the suggestions 
to mitigate the possibility of tag explosion and tangled topics, 
as well as to ensure the semantic uniqueness of the relation types.
Once a suggestion is made, it must be validated 
by at least three other contributors 
before it is integrated into the KG.
The refined relationships and topics 
amassed through the Acquisition phase
will be the ones subject to evaluation in the next snapshot.
Later, we added automation measures and restriction policies 
into the evolved version of the online platform 
to minimize the manual work required of the maintainers.

\subsection{Phase 2: Automated Topic Recommendation}
In this phase, 
we propose two topic prediction models 
for two different scenarios; 
topic augmentation and topic set recommendation.
We build these recommenders
using the semantic information obtained 
in the previous phase.

\subsubsection{Topic Augmentation}\label{sec:recom_approach_pure}
\begin{figure}[bt!]
    \centering
    \includegraphics[width=0.35\linewidth]{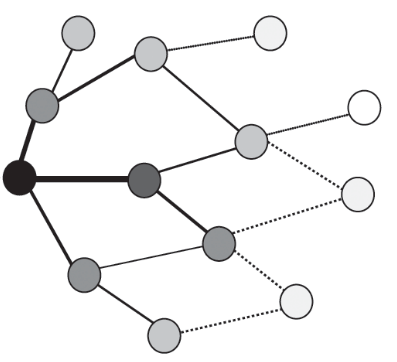}
    \caption{Spreading Activation}
    \label{fig:recom_approach_san}
\end{figure}

We first propose KGRec, 
a model that takes the topics 
already assigned to a software project as input 
and recommends \textit{missing} but relevant topics.
Note that the input to this model is 
only the seed set topics and no extra information about the repository.
Using the SED-KGraph we aim to expand this initial set.
We apply a \textit{spreading activation} technique~\cite{crestani1997application} 
on SED-KGraph to depth one.
Spreading activation techniques 
operate on semantic networks 
based on the Spreading Activation theory in semantic networks.
\autoref{fig:recom_approach_san} 
displays an overview of the spreading activation effect. 
When a set of nodes are activated, in the graph
spreading activation computes the activation score of other nodes.

We first annotate SED-KGraph with node weights
computed based on the popularity of the topic in the community 
and the degree of the topic node in SED-KGraph.
Our intuitions are; 
(1) if a topic is used frequently in the community, 
its a more useful and valuable topic,
and (2) a topic with a higher degree in SED-KGraph 
is related to more topics, 
and consequently has a better chance of being relevant 
to more projects which might be assigned the related topic.
The weights are calculated as
\begin{equation}
    W_t = \alpha \times P_t + \beta \times D_t,
    \alpha + \beta = 1
\end{equation}
in which $W_t$ denotes the weight of the node corresponding to topic $t$. 
$P_t$ and $D_t$ are defined as below 
as a measure of popularity and degree of 
the topic $t$, respectively
\begin{equation}
    \begin{split}
        & P_t = \frac{\log{(n_t + 1)}}{\max\limits_{t_i \in T} \log{(n_{t_i} + 1)}}\\
        & D_t = \frac{\log{(d_t + 1)}}{\max\limits_{t_i \in T} \log{(d_{t_i} + 1)}}
    \end{split}
\end{equation}
where $n_t$ is the number of projects 
in the platform labeled with topic $t$, 
$d_t$ is the total degree of topic $t$ in SED-KGraph, 
and $T$ is the set of topics in SED-KGraph.
Note that $ \alpha $ and $ \beta $ 
are coefficients to scale the scores.
Note that the results 
can vary for different values of $\alpha$ and $\beta$. 
For instance, for $\alpha > \beta$, 
the results will mostly rely on 
the set of most popular GitHub-featured topics.  
With  $\alpha < \beta$, 
one can emphasize topics 
that are not widely used by the community yet. 
To account equally for both popularity and node degree, 
we set $ \alpha = \beta = 0.5$.
Future researchers and practitioners 
can set these values according 
to their needs and use cases.

To generalize the formulation of the approach 
such that it can apply to the KGRec+ model as well, 
consider that topic $t$ is relevant to project $p$ 
with the probability ${Pr}_t^p$ (here ${Pr}_t^p = 1$ 
since topics are already assigned to the projects).
$I$ is the set of initial topics assigned to project $p$ 
and $N(I)$ denotes the set of topics 
that are immediate neighbors to at least one of the topics in $I$. 
For all $t$ in $N(I)$,
we spread this probability 
to compute the relevance score of topic $t$ to project $p$ 
along SED-KGraph edges using Equation~\ref{eq:spread}.
Then, the model sorts the list of candidate topics 
by the $S_t^p$s and return the top-$k$ ones 
as a list of recommendations.
\begin{equation}\label{eq:spread}
    S_t^p = W_t \times \sum\limits_{t_k \in I} {Pr}_{t_k}^p ; \forall t \in N(I)
\end{equation}

Note that in our current model, 
we do not take the relation type into account 
when spreading the activation along the edges of the KG. 
Doing so requires extensive analysis of whether certain types of relations 
result in augmented topics that are a better fit for the repository. 
As the evaluation of the recommendations 
is a human-intensive task, 
verifying the effect of the relation types 
on the quality of the recommendation 
can only be achieved in the long run 
when the recommender system has been used enough times 
to provide adequate data for the analysis. 
We leave this as a future direction for this research.

\subsubsection{Topic Set Recommendation}\label{recom_approach_ml}
In this application, 
we assume that repositories are not labeled with any initial topic.
We build upon KGRec and propose KGRec+ 
as a stand-alone topic set recommender model.
We feed the model with the available textual data on the projects.
This textual data include README files, repository description, and wiki pages.
We concatenate and then transform these pieces of textual data
to their respective TF-IDF vectors 
for consumption by the classifiers.
Note that we use the \textit{preprocessed} dataset 
provided by Izadi et al.~\cite{izadi2021topic} 
in their recent work on the topic recommendation.
\footnote{\url{https://github.com/MalihehIzadi/SoftwareTagRecommender}}
After the model is trained on repositories' textual data and assigned topics,
it can predict a list of relevant topics 
for a given project.
Finally, we take this set predicted by the ML-based component 
and complement it using KGRec.

As the ML-based component in our proposed model, 
we use two classifiers 
employed in our baselines. 
Based on Di Sipio et al.'s approach, 
we train a \textit{Multinomial Naive Bayes} (MNB) text classifier 
that takes as input the textual data 
on the software project and predicts the ${Pr}_t^p$s 
for the GitHub featured topics.
We also take Izadi et al.'s approach 
and train a multi-class multi-label \textit{Logistic Regression} (LR) classifier 
which operates similarly to the baseline 
and yet significantly outperforms it~\cite{izadi2021topic}.
Note that we follow the instructions 
provided by the baselines 
to build these classifiers.

The ML-based component, trained with textual data of projects, 
predicts the ${Pr}_t^p$ for each topic.
We take the top-$m$ topics with the highest ${Pr}_t^p$ 
and feed them to the KGRec component as the set $I$.
The KGRec component operates on this set of topics 
as input and augments the list of $m$ topics 
with $g$ more topics to return a list of $k = m + g$ topics as the recommendations.
We propose LR+KGRec as our main approach and show that it outperforms all models in terms of all the considered metrics.

\section{Experiment Settings}\label{sec:settings}
In this section, we present our experimental setting.
We first state the research questions of this study,
then review the dataset, evaluation metrics, and model setting.
Next, we provide an overview of our baselines, 
contributors' information engaged in the construction of SED-KGRaph,
and the platform using which the graph was built.

\subsection{Research Questions}\label{sec:rq}
We aim to augment topic recommender models for software projects semantically
using a high-quality KG tailored to SE topics.
Thus, we answer the following research questions:
\begin{itemize}
    \item \textbf{RQ1: 
    What are the characteristics of 
    our collaboratively-constructed KG, SED-KGraph?}
    First, as the dataset containing the KG 
    is one of our main contributions, 
    we describe the KG characteristics 
    including its size, entities quantities or types, 
    contributors' agreement rates during each snapshot, and more. 
    As with any other dataset contributed to 
    the research or industry community, 
    we aim to report these characteristics 
    to clarify details for the readers and future users of the KG. 
    Moreover, as the KG is susceptible 
    to expand over time, 
    we would like to report our contribution 
    at the time of conducting the research.
    \item \textbf{RQ2: 
    How accurately can we 
    augment a set of initial topics assigned to a repository utilizing SED-KGraph?}
    Missing topics recommendation is one of the main applications of SED-KGraph. 
    Hence, with RQ2 we assess the accuracy of recommenders 
    which solely augment missing topic sets using a seed of initial topics.
    \item \textbf{RQ3: 
    How accurate is our topic set recommender, KGRec+, 
    as a stand-alone predictive model?}
    Finally, we would like to build upon the previous application
    and propose a stand-alone recommender 
    that takes a repository's textual information and predicts relevant topics. 
    This is different from the previous question 
    in which we only feed the model the initial topic seeds 
    and not the repository's textual information.
\end{itemize}

With these three questions, 
we hope to clarify our contribution regarding the KG itself
and also the two semantically-enhanced predictive models built upon such a graph.

\subsection{Dataset}\label{sec:recom_experiment_data}
We use the dataset from Izadi et al.'s study~\cite{izadi2021topic}\footnote{\url{https://github.com/MalihehIzadi/SoftwareTagRecommender}}, 
which contains cleaned textual data 
on about $152$K projects, 
along with their cleaned topics.
Our set of topics in the dataset contains $236$ 
of GitHub's featured topics,
eight more topics 
compared to the baseline~\cite{izadi2021topic}.
This small difference (eight topics) 
comes from a mapping of synonymous topics in the dataset 
to their corresponding GitHub-featured ones 
based on the golden mapping provided by the baseline. 
This resulted in a strengthened number of samples 
for the few topics which qualified them
for the training of the ML-based component.
Note that all repositories have owner-assigned topics.
In this dataset, 
all repositories have at least one topic,
and on average, 
$2.46$ topics are assigned 
to each repository by their owners.
The textual data we use to train the models include 
repositories' descriptions, README files, and wiki pages.
We do not use source file names and project names.
In the provided preprocessed dataset~\cite{izadi2021topic}, 
about $12\%$ of repositories have wiki pages.
Considering that all repositories in the data have at least one README file, 
we have ample data even for those that do not have wiki pages. 
Hence, we do not perform additional preprocessing tasks on the data.
The baseline study provides 
more details on the statistics of data~\cite{izadi2021topic}.
Similar to the common standard, 
we take $80\%$ of the $152$K projects 
as the training set for our ML-based component 
and leave the remaining $20\%$ as the test set.

\subsection{Evaluation Metrics}\label{sed:kg_experiment_eval}
We report the \textit{Success Rate} (SR) of the relationships 
in the partially objective labeling task 
as a measure of the quality of the relationships 
defined by the maintainers during KG Initialization 
or suggested by the contributors during Acquisition.
SR is the ratio of the relationships labeled ``True" 
over the set of all the relationships under evaluation.
The relationships labeled with ``True" are considered successfully defined.
Thus, with $N_T$ as the number of relations labeled as ``True" 
and $N_F$ as the number of relations labeled as ``False", the metric is defined as
\begin{equation}
    SR = \frac{N_T}{(N_T+N_F)}.
\end{equation}

We also report the \textit{Absolute Agreement Ratio over True Relationships} (AARTR)
as a measure of the quality of the community-approved relationships.
A relationship is absolutely agreed upon 
if all the contributors label it with the same label (``True" or ``False").
Respectively, a true relationship is absolutely agreed upon 
if all the contributors label it as ``True".
Considering $N_T^{AA}$ as the number of absolutely agreed true relationships 
and $N_T$ as the total number of true relationships, we define AARTR as
\begin{equation}
    {AARTR} = \frac{N_T^{AA}}{N_T}.
\end{equation}

Finally, to quantify the reliability of the contributors 
in the Evaluation step, 
we report \textit{Average Rater-Objective Conformance Rate} (AROCR).
\textit{Rater-Objective Conformance Rate} (ROCR) 
is how the reliability of raters (contributors) 
is measured in a crowd-sourced partially objective labeling task~\cite{alonso2014crowdsourcing}.
Considering $N_{R_i}^C$ as the number of votes from $R_i$ 
that conform with the final objective label of the items (relationships) 
and $N_{R_i}$ as the total number of votes from $R_i$, the ROCR for rater $R_i$ is calculated as
\begin{equation}
    {ROCR_{R_i}} = \frac{N_{R_i}^C}{N_{R_i}},
\end{equation}

AROCR for the labeling task is defined 
as the average of $ROCR_{R_i}$ over all raters who contributed to the task.
We select this metric since we engage a large number of contributors.
Moreover, not all of our contributors review all the relationships, 
which makes metrics such as Cohen's Kappa an inadequate measure of reliability for our case.

Please note that Cohen's Kappa coefficient 
is a statistic that is used 
to measure inter-rater reliability for qualitative items. 
However, in this study, we have \textit{multiple} and \textit{varying} 
raters per relationship, 
hence, we cannot incorporate Cohen's Kappa. 
Moreover, in most cases, our raters share only a few relationships 
they have voted for with each of the other contributors. 
This incurs a huge amount of missing data 
and makes the results even more sensitive to human error in votes. 
As previous crowd-sourced studies have also exemplified~\cite{zhang2021method}, 
even Krippendorf's alpha which accounts for missing data 
is not a well-suited metric for crowd-sourced studies 
with large numbers of participants. 
Hence, we used AARTR.

To evaluate the recommender systems, 
we run all approaches on the test set ($20\%$ of the projects in the dataset).
However, since KGRec recommends ``missing" topics, 
both as a stand-alone model and as a component of KGRec+, 
automatically calculated classification metrics 
such as Recall and Precision 
are irrelevant to the purpose of this study~\cite{izadi2022evaluation}.
We support this statement by evaluating two baseline approaches~\cite{di2020multinomial,izadi2021topic} 
both automatically against the ground truth 
and manually through human evaluation, 
and list the results in \autoref{tab:auto_vs_man}.
The results show that since the ground truth 
does not contain the missing topics, 
the automated evaluation of the recommendation lists does not serve justice to the merits of each model in recommending missing topics.
Therefore, we randomly sample $50$ projects 
from the test set and evaluate the results 
through a human evaluation process with five experts.
Three evaluators validate the results for each project.
We engage five Computer Science experts 
from both academia and industry 
in this human evaluation process.
For each sampled project, we ask three evaluators 
to go through the content of the project
and then determine whether the recommended topics were relevant to the project.
To avoid biasing our evaluators towards any one of the approaches, 
we shuffle the results from different approaches 
before anonymously presenting them to the evaluators.
KGRec is evaluated against TopFilter, 
and the ML+KGRec is evaluated against the ML+TopFilter 
as their goals are aligned.
We also include evaluations on the ML-based components 
to investigate how KGRec can improve a ML-based solution.
\begin{table}[!tb]
\centering
\caption{Automated versus manual evaluation}
\label{tab:auto_vs_man}
\begin{tabular}{|l|c|c|}
\hline
\multirow{2}{*}{\textbf{Model}} &
\multicolumn{2}{c|}{\textbf{ASR@5 Evaluation Method}} \\ \cline{2-3}
& \textbf{Automated} & \textbf{Manual} \\ \hline
Di Sipio et al.~\cite{di2020multinomial} & 24.60\% & \textbf{30.80}\% \\\hline
Izadi et al.~\cite{izadi2021topic} & 30.00\% & \textbf{50.80\%} \\\hline
\end{tabular}
\end{table}

As a measure of the practicality of the missing topic recommendation task, 
we report the percentage of the test cases in the test set 
and in the test sample set for which each of the approaches 
fails to recommend any topics.
For this purpose, with $N_{FC}$ as the number of test cases a model 
fails to return any recommendations for
and $N_C$ as the total number of test cases in the set, 
we define \textit{Failed Case Ratio} (FCR) as
\begin{equation}
    {FCR} = \frac{N_{FC}}{N_C}.
\end{equation}

To quantify the quality of the recommendations,
we report \textit{Average Success Rate $@k$} (ASR$@k$) 
to evaluate the performance of the topic recommendation approaches.
We also report \textit{Mean Average Precision $@k$} (MAP$@K$) metric,
a commonly-used measure for evaluating recommender systems 
that returns a ranked list of results.
It captures how high successful suggestions are positioned in the recommendation list as well as the number of successful suggestions.
Note that as FCR is already reported in the missing topic recommendation task's results, we only report the ASR and MAP over the set of test case samples for which the KGRec or TopFilter components do return a list of recommendations.
However, reporting FCR for the automated topic recommendation task is irrelevant since the recommendation list is never empty.

\subsection{Model Settings}\label{sec:model_setting}
First, to calculate the topic weights ($W_t$), 
we get the number of public repositories 
labeled with each of the $863$ topics in SED-KGraph through GitHub API calls.
We then construct the weighted graph and implement KGRec in a tailored Python script.
As we use the dataset from the baseline study~\cite{izadi2021topic}, 
the mean and maximum number of tokens 
in the concatenated input data is $235$ and $650$ tokens. 
Similar to the baseline paper, 
we set the maximum input length to $512$ tokens 
and the maximum number of features 
to $20K$ for TF-IDF embedding vectors.
The input data to the ML-based components 
is repositories' descriptions, README files, and wiki pages.
We train the MNB~\cite{di2020multinomial} and LR~\cite{izadi2021topic} models from the python library, \texttt{scikit-learn}, for $236$ of the GitHub featured topics, 
as these are the topics with enough supporting instances for training in the dataset.
We set $k$ to $5$ as the length of recommendation lists.
When evaluating the KGRec+ models, 
as the average number of topics assigned to the projects is $2.46$, 
we set $m$ to three.
Then, feeding these $m$ topics to the KGRec component, 
we take top-$f = 2$ topics from KGRec 
to make a full list of top-$5$ recommendations.

\subsection{Baselines}
\paragraph{Baselines for Topic Augmentation}
TopFilter, the state-of-the-art approach 
for augmenting a repository's topics list based on its initial set,
is an item-based collaborative filtering approach~\cite{dirocco2020topfilter}.
In this approach, each project is represented 
with a set of assigned topics using a project-topic matrix.
For each project, taking the set of topics assigned to the project, 
the model computes the similarity of this topic set with the topic sets of all other projects in the dataset and takes the topics assigned to the top-$25$ most similar projects (in terms of the set of topics assigned to the project) 
as the candidate set of topics.
Calculating a ranking metric defined by the authors, the model returns the top-$k$ topics as recommendations.
While the authors do not evaluate TopFilter as a single component, 
we use this model as the baseline for our first task (topic augmentation).

\paragraph{Baselines for Topic Set Recommendation.}
For the second task, 
we use \textit{MNB+TopFilter} proposed by 
Di Rocco et al.'s~\cite{dirocco2020topfilter} 
as one of the baselines.
We also compare our method against Izadi et al.'s~\cite{izadi2021topic} 
and Di Scipio et al.'s~\cite{di2020multinomial} 
proposed methods for the topic recommendation.
The third baseline uses an LR classifier based on our previous work.  
LR takes a repository's textual information
including README files, description, and available wiki pages,
concatenate them, and transforms them into TF-IDF vectors. 
Then the classifier is trained on the TF-IDF vectors. 
The labels for the classifier are the assigned topics.
To provide a more comprehensive evaluation, 
we also stack our model (KGRec) on 
Di Scipio et al.'s~\cite{di2020multinomial} proposed model.
Moreover, we combine Di Rocco et al.'s~\cite{dirocco2020topfilter} 
and Izadi et al.'s~\cite{izadi2021topic} approaches together
and introduce it as another baseline.

\subsection{Platform}\label{sec:platform}
We designed and deployed an online platform  
to collect contributions 
from the SE community and use their help 
in building and evaluating topics, relation types, and relationships 
(collectively called KG entities). 
First to construct the KG, 
we automatically retrieved GitHub's featured topics 
and presented them to contributors 
for knowledge acquisition in our platform.
Contributors performed 
CRUD operations on KG entities independently.
Then, to minimize human effort and facilitate 
the KG Expansion step, 
we added more functionalities 
to the online platform.
In the following, 
we describe our platform 
and it features with more details.
Generally, after logging in, contributors can access 
and modify their previous contributions 
through their 
dashboard.~\footnote{To access the platform, please refer to our public GitHub repository at \url{https://github.com/mahtab-nejati/KGRec}.}
For snapshots of the platform 
please refer to the appendix.

\paragraph{Platform Functionalities}
Inspired by similar crowed-sourced projects 
in code hosting and information websites 
such as GitHub and Stack overflow
we provide a set of functionalities 
to help maintain the KG as it grows.
These functionalities provide 
a level of autonomy for contributors 
while keeping the aforementioned challenges under check.
They are used/performed 
by contributors, maintainers, or the platform itself.
More specifically, users can contribute to the expansion 
and maintenance of the KG in several ways, including
\begin{itemize}
\item vote to approve or disapprove an already defined relationship,
\item suggest new relationships for each of the topics they review through free-text forms,
\item introduce new KG entity,  
\item edit their previously suggested KG entity,
\item remove their own suggestion (before they are featured on the platform),
\item request edit for existing KG entities 
so that SED-KGraph is modifiable 
upon evolution and emergence of topics,
\item report spams, and finally 
\item report duplicates or aliases manually. 
Such reports are then brought 
to the entity owners' and maintainers' attention to address.
This solution can also help with semantically-similar 
topics or relationships that are written with different lexicons.
\end{itemize}

The platform automatically runs the following actions.
\begin{itemize}
\item verify new entities based on our policies 
described in the following,
\item run reliability check for contributors,
\item check for possible aliases/redundancies. 
This is done automatically to check whether the topics 
that are being introduced by the contributors 
are already defined in the KG. 
We detect aliases and redundancies 
based on the topics' names, and alias lists.
Specifically, we use the NLTK library's edit distance functionality.\footnote{\url{https://tedboy.github.io/nlps/generated/generated/nltk.edit_distance.html}}
Through this mechanism, 
the edit distance between all names (full name and display name) 
and aliases of each pair of topics are calculated. 
Thus, for a pair of topics $t_1$ and $t_2$ 
with $m_1$ and $m_2$ names 
(full name, display name, and aliases) respectively, 
$m_1 \times m_2$ 
similarities are computed. 
The pair is marked as potential redundancy 
if at least one of these similarities 
is above a certain threshold, here $80\%$. 
We choose this threshold 
to retrieve all potential duplicates 
and minimize the False Negative error.
However, this value can be tuned.
Once a topic is detected as potentially duplicate, 
it is listed along with the pair causing the duplicate 
for maintainers to check on them.
\end{itemize}

Finally, maintainers perform the below tasks to keep the KG intact.
The platform automatically runs the following actions.
\begin{itemize}
\item random checks,
\item verify edits, 
\item check reported spams, and
\item resolve reported redundancies. 
The potential redundant pairs identified
either manually by contributors or automatically by the system are  
brought to the entity owners' and maintainers' attention to address.
\end{itemize}

\paragraph{Platform Policies}
We also establish policies to guarantee contributors' eligibility 
for evaluation and expansion of SED-KGraph.
The policies include tutorials before granting permission 
to perform CRUD operations, 
random checks on the suggestions and evaluations by maintainers,
and reliability checks based on the conformance of contributors' answers 
with the objective labels of relationships.
This helps identify issues in the KG 
and potentially detecting unreliable users.
More specifically, 
we have different policies in place for 
user permissions and entity acceptance 
as explained in the following.

1. User Permission Policies
\begin{itemize}
    \item Only users with at least three years 
    of academic experience or one year of industrial experience 
    and Computer Science-related fields 
    shall contribute to the study. 
    Once a user meets the minimum requirement, 
    they are considered reliable to start contributing. 
    \item We constantly check for the reliability of the users 
    by comparing whether the majority of their votes 
    conform with the majority agreement for each relationship 
    they have voted for. 
    If the portion of their conforming votes 
    falls under a set threshold, 
    here 50\%, 
    their reliability is revoked 
    and all their previous votes are nullified. 
    Note that this threshold can be configured to other values as well. 
    \item We provide background information on topics (if any). 
    For instance, several topics 
    already have definitions in the GitHub featured set.
    \item Reliable contributors can up/down-vote the relationships. 
    To vote for a relationship, 
    contributors need to first read the definition 
    of the verb in the relationship and mark it as read.
   \item Contributors are also allowed 
    to skip assessing a relationship in case 
    they do not have enough knowledge 
    to evaluate it or simply prefer not to comment. 
    \item Reliable contributors get creator-level permissions 
    if they have marked all the verbs as read 
    and voted for a total of $50$ relationships involving $20$ topics.
    Similarly, the parameter values here can also be tuned.
    As a creator, contributors can define new topics, verbs, and relationships.
    \item Take note that once the reliability is revoked, 
    creator-level permissions are revoked too.
    \item Contributors can edit or delete the entities they have created 
    unless the entity is accepted (featured in the platform).
\end{itemize}

2. Entity Acceptance Policies
\begin{itemize}
\item Relationships are accepted
through a partially objective labeling task.
\item In this task, contributors up-vote or down-vote each of the relationships. 
They can also declare that they do not know whether the relationship is correct, 
in which case the vote is considered a null vote 
and does not affect the acceptance criteria.
\item For a relationship to be accepted, 
at least three non-null votes 
from the contributors are required.
\item If all first three voters agree 
that a relationship is correct (absolute agreement, $100\%$ acceptance), 
the relationship is accepted and featured.
Otherwise, we gradually lower the threshold for acceptance down to $65\%$ 
among at least nine contributors. 
We do not lower the threshold any further
to make sure that suspicious relationships are not accepted. 
Note that we tried lowering it down to $50\%$ 
and the graph did not change a lot. 
However, we chose not to decrease the threshold 
to have higher confidence in the result.
\item A topic/verb is accepted and featured 
if it is in at least one accepted relationship.
\item GitHub-featured topics are also accepted 
by default as they have already been assessed 
by the GitHub community and the project's maintainers.
\end{itemize}

\subsection{Contributors' Overview}
\label{sec:kg_experiment_participants}

In this section, we provide an overview of contributors.
To engage contributors and ensure diversity, 
we sent out invitations to the technical teams of $30$ local technology-based companies active in a variety of SED-related fields 
including software engineering, cloud computing, data science, network, blockchain, security, social media, e-commerce, digital advertisement, entertainment, etc.
We also invited students of related programs including Computer Engineering, Computer Science, Data Science, Software Engineering, IT, etc. 
from $20$ top local and international universities.
The invitation was open for a total of seven months, 
over which individuals could apply to participate in the study.

To improve reliability, we disqualified 
(1) students with less than three years of academic experience and no industrial experience,
and (2) practitioners with less than three years of industrial experience who did not have at least three years of prior academic experience.
Since the first six months of KG Expansion 
required central control over the suggestions, 
the maintainers thoroughly refined the suggestions 
every two months and issued refined suggestions for evaluation.
Therefore, the first three snapshots of the SED-KGraph were captured.
We engaged the first $50$ applicants
during the first, the next $40$ applicants
in the second, and the next $30$ applicants
in the third two-months period of KG Expansion, 
resulting in the three snapshots.
Finally, the last $50$ applicants were engaged 
in a long-term (six months) snapshot 
for expansion of the KG (fourth snapshot).
Throughout the KG Expansion step, 
we eliminated and replaced unreliable contributors, 
i.e., contributors with \textit{Rater-Objective Conformance Rate} 
(ROCR, defined in Section~\ref{sec:kg_approach_expansion_evaluation})
lower than $50\%$.
The reliability checker functionality of the online platform 
automatically applies this policy among others to assure the reliability of the contributors.

In the end, $170$ individuals have made contributions 
to the study from $16$ companies and $11$ universities.
The diversity of the contributors' experience and expertise matched our requirements, considering the wide range of topics in the KG, 
and allowed for a fair evaluation and expansion process.
\autoref{tab:participants} presents more details, 
including the average years of experience in both industry and academia, 
on the contributors.
\begin{table}[tb]
\centering
\caption{Overview of contributors' information}
\label{tab:participants}
\begin{tabular}{c|c|c|c|cc|ccc|ccc}  \hline
& & & & \multicolumn{2}{c|}{Gender} 
& \multicolumn{3}{c|}{\pbox{25mm}{Experience\\(Academia, years)}} 
& \multicolumn{3}{c}{\pbox{25mm}{Experience\\(Industry, years)}} \\ \cline{5-12} 
\multirow{-3}{*}{\rotatebox{40}{Snapshot}} & 
\multirow{-3}{*}{\rotatebox{40}{Duration}} & 
\multirow{-2}{*}{BG} & 
\multirow{-2}{*}{All} 
& M & F 
& Avg & Min & Max 
& Avg & Min & Max \\ 
\midrule

\multirow{3}*{\textbf{\#1}} & \multirow{3}*{2M} 
& Academia & 25 & 15 & 10 & 3.24 & 3 & 5 & 0 & 0 & 0 \\\cmidrule{3-12}
& & Industry & 25 & 21 & 4 & 3.92 & 3 & 10 & 1.92 & 1 & 9 \\\cmidrule{3-12}
\rowcolor{beaublue} & &  All & 50 & 36 & 14 & 3.58 & 3 & 10 & 0.96 & 0 & 9 \\ 
\midrule

\multirow{3}*{\textbf{\#2}} & \multirow{3}*{2M} 
& Academia & 13 & 10 & 3 & 5.08 & 3 & 10 & 0 & 0 & 0 \\\cmidrule{3-12}
& & Industry & 27 & 13 & 14 & 7.81 & 0 & 16 & 5.85 & 1 & 10 \\\cmidrule{3-12}
\rowcolor{beaublue} & &  All & 40 & 23 & 17 & 6.73 & 0 & 16 & 3.95 & 0 & 10 \\ 
\midrule

\multirow{3}*{\textbf{\#3}} & \multirow{3}*{2M} 
& Academia & 12 & 5 & 7 & 5.83 & 4 & 10 & 0 & 0 & 0 \\\cmidrule{3-12}
& & Industry & 18 & 17 & 1 & 7.94 & 4 & 15 & 4.56 & 2 & 14 \\\cmidrule{3-12}
\rowcolor{beaublue} & &  All & 30 & 22 & 8 & 7.1 & 4 & 15 & 2.74 & 0 & 14 \\ 
\midrule

\multirow{3}*{\textbf{\#4}} & \multirow{3}*{6M} 
& Academia & 18 & 7 & 11 & 4.21 & 3 & 9 & 0 & 0 & 0 \\\cmidrule{3-12}
& & Industry & 32 & 15 & 17 & 5.46 & 2 & 8 & 3.11 & 1 & 6 \\\cmidrule{3-12}
\rowcolor{beaublue} & & All & 50 & 22 & 28 & 5.01 & 2 & 9 & 1.99 & 0 & 6 \\ 
\bottomrule
\end{tabular}
\end{table}

\section{Results}\label{sec:resutls}
In this section, we provided the results of our approach.
We first review the characteristics of 
the constructed KG to answer RQ1.
Then, we proceed to present 
the evaluation results of the two recommender models 
to address RQ2 and RQ3.

\subsection{SED-KGraph: Data Characteristics}\label{sec:kg_resutls}
This section first summarizes 
the results of each step in the KG construction process, 
followed by the characteristics of SED-KGraph.
\autoref{tab:kg_stats} and \ref{tab:kg_results} 
summarize the results captured in each snapshot.

\begin{table*}[tb]
\centering
\caption{Statistics for different snapshots.}
\label{tab:kg_stats}
\begin{tabular}{c|ccc}
\toprule
Snapshot & Topics & Relationship Types & Verified Relationships 
\\ \midrule

\#1 & 461 & 4 & 0 \\ \midrule
\#2 & 640 & 12 & 982 \\\midrule
\#3 & 716 & 13 & \numprint{1548} \\\midrule
\#4 & 812 & 13 & \numprint{1864} \\ \midrule
All & 863 & 13 & \numprint{2234} \\ \bottomrule
\end{tabular}
\end{table*}
\begin{table*}[tb]
\centering
\caption{Results of expansion and maintenance tasks on the graph}. 
{\small (*) TL and FL denote the number of True Labels and False Labels.}
{\small (**) These topics and relationships are gradually added to and evaluated in the snapshot\#4.}
\label{tab:kg_results}
\begin{tabular}{c|cc|ccc|ccc} 
\toprule
& \multicolumn{5}{c}{Evaluation (Relationships)} 
& \multicolumn{3}{c}{Acquisition (Suggestions)} \\ 
\cmidrule{2-9}
Snapshot & TL & FL &
\cellcolor{beaublue}{SR} &
\cellcolor{beaublue}{AARTR} &
\cellcolor{beaublue}{AROCR} &
\pbox{10cm}{Relationship Types} & Topics & Rels \\ 
\midrule

\#1  
& 982 & 101 &
\cellcolor{beaublue}{0.907} &
\cellcolor{beaublue}{0.887} &
\cellcolor{beaublue}{0.791} &
8 & 179 & 635 \\\midrule

\#2
& 566 & 69 &
\cellcolor{beaublue}{0.891} &
\cellcolor{beaublue}{0.744} &
\cellcolor{beaublue}{0.726} &
1 & 76 & 322 \\\midrule

\#3 
& 316 & 6 &
\cellcolor{beaublue}{0.981} &
\cellcolor{beaublue}{0.877} &
\cellcolor{beaublue}{0.891} &
0 & 15 & 39 \\\midrule

\#4 
& 370 & 41 &
\cellcolor{beaublue}{0.900} &
\cellcolor{beaublue}{0.754} &
\cellcolor{beaublue}{0.780} &
0 & 51** & 372** \\ \midrule

\rowcolor{beaublue} 
All 
& \numprint{2234} & 217 &
\cellcolor{beaublue}{0.911} &
\cellcolor{beaublue}{0.828} &
\cellcolor{beaublue}{0.789} &
9 & 321 & 1368 \\ 
\bottomrule
\end{tabular}
\end{table*}

\paragraph{Initialization}
\label{sec:kg_results_initialization}
The seed set of topics from GitHub's project 
contained $389$ topics from a wide variety of areas in SED, 
and from different levels of abstraction, 
i.e., topics could be as coarse-grained as \texttt{ai} 
or as fine-grained as \texttt{django}.
However, this set proved to be relatively inadequate 
in representing the final goal of the study 
since many proper topics 
such as \texttt{web-development}, and \texttt{ui-ux} were missing.
Moreover, such an occurrence is inevitable 
due to the constant emergence of new topics in the community.
The maintainers identified this issue 
while defining the relationships and as a solution, 
augmented the seed set with $72$ more topics 
in the process of constructing the first draft of SED-KGraph, 
enlarging the set to include $461$ distinct topics.
This was done with the aim 
to provide a more comprehensive set 
and to facilitate the contributors' task.
Having conciseness in mind, 
maintainers defined four primary relation types, 
namely \textit{is-a}, \textit{is-used-in-field}, 
\textit{provides-functionality}, and \textit{works-with}
described and exemplified 
in \autoref{tab:relation_type}.~\footnote{For more samples please refer to Appendix~\ref{appendix:sample_rels}}
The first three relation types 
capture three determinative characteristics of a topic 
regarding the topic's scope.
The last relation type 
connects the most closely intertwined 
yet differently categorized topics together.
In the end, the maintainers agreed on \numprint{995} relationships 
and disagreed over a total of \numprint{88} of them ($8.13\%$).
This yielded \numprint{1083} relationships 
with the four primary types among the \numprint{389} featured topics 
and the $72$ augmented topics in the initial draft of SED-KGraph.

\begin{figure*}[tb!]
    \centering
    \begin{subfigure}[b]{0.48\textwidth}
        \centering
        \includegraphics[width=\linewidth]{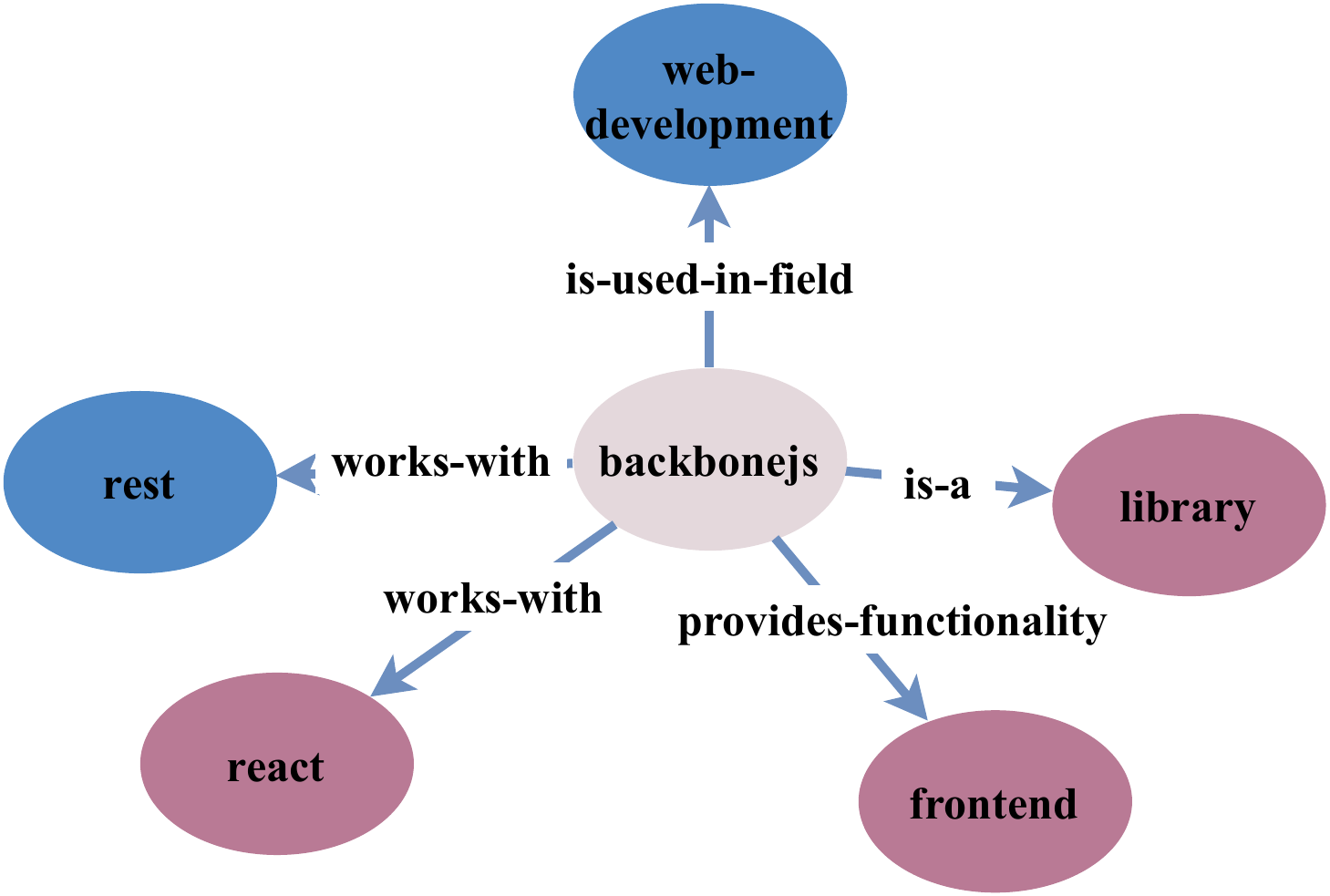}
        \caption{The First Snapshot}
        \label{fig:sample_1}
    \end{subfigure}
    \begin{subfigure}[b]{0.48\textwidth}
        \centering
        \includegraphics[width=\linewidth]{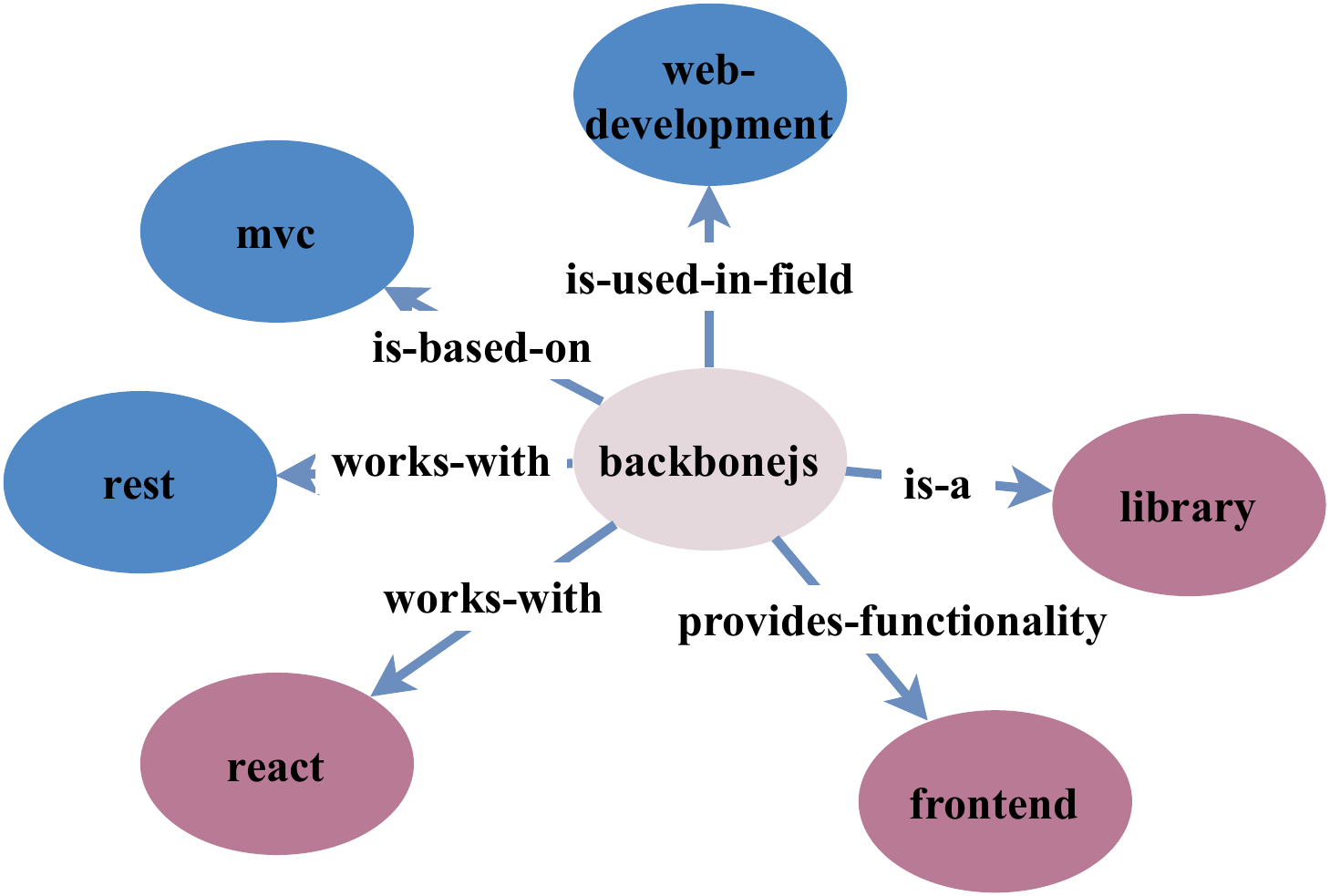}
        \caption{The Second Snapshot}
        \label{fig:sample_2}
    \end{subfigure}
    \begin{subfigure}[b]{0.48\textwidth}
        \centering
        \includegraphics[width=\linewidth]{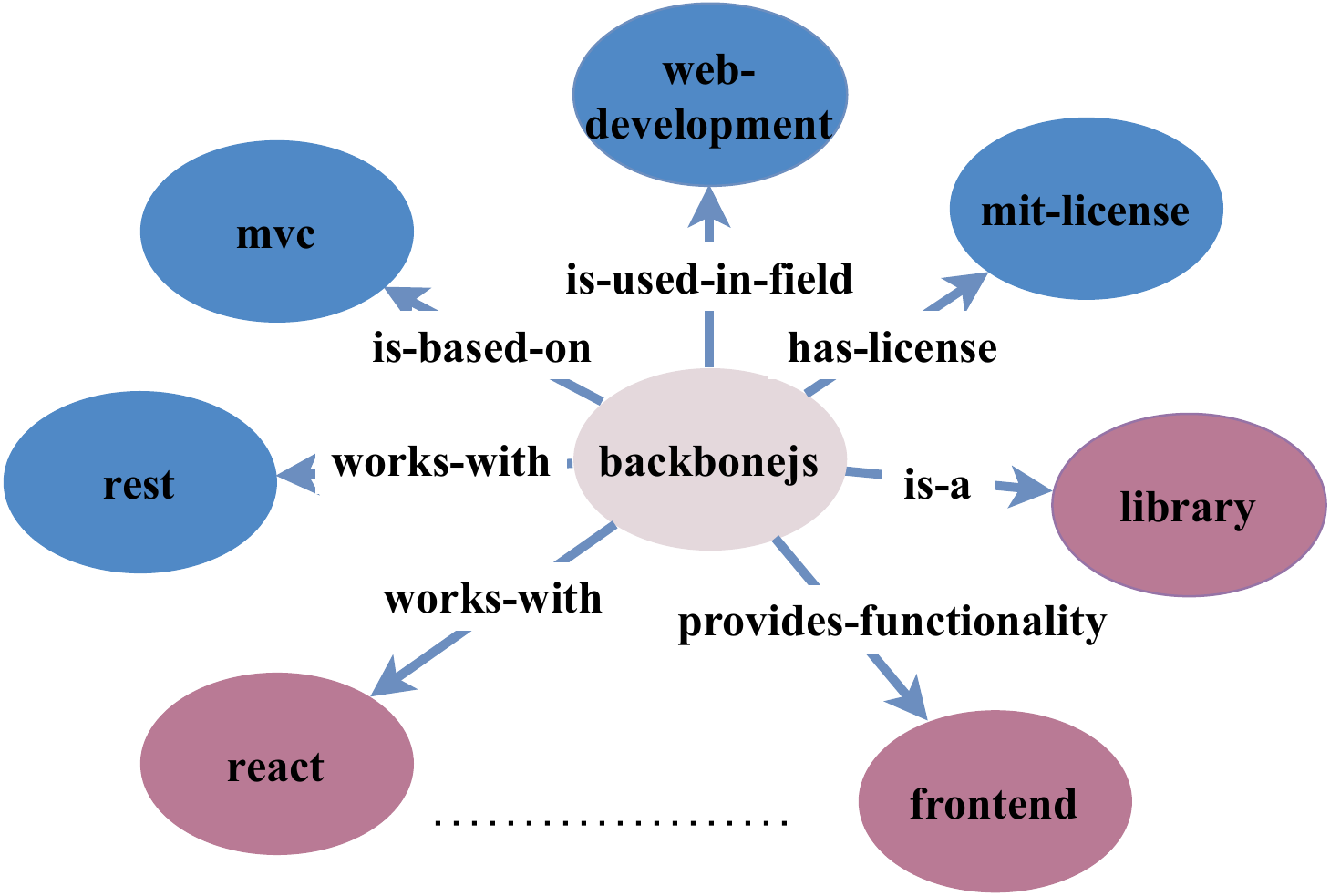}
        \caption{The Third Snapshot}
        \label{fig:sample_3}
    \end{subfigure}
    \begin{subfigure}[b]{0.48\textwidth}
        \centering
        \includegraphics[width=\linewidth]{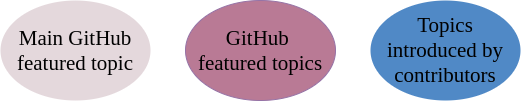}
        \caption{Legend}
        \label{fig:sample_l}
    \end{subfigure}
    
    \caption{Sample Node Expansion.}
    \label{fig:smaple}
\end{figure*}

\paragraph{Expansion}\label{sec:kg_results_expansion}
Figure~\ref{fig:smaple} illustrates a sample node
and its relationships' evolution over 
the KG Expansion process in the SED-KGraph.

In the first snapshot, 
we evaluated the initial draft of SED-KGraph.
From the $995$ relationships in the initial KG, 
(excluding the $88$ that the maintainers disagreed on) 
contributors labeled $963$ as ``True"
but disapproved $32$ relationships.
Contributors also rejected $69$ of relationships 
already disapproved by the maintainers 
and accepted only $19$ of them.
This yielded a success rate of $96.78\%$ for the set of $995$ relationships, 
$21.60\%$ for the set of $88$ relationships, and $90.67\%$ in total.
Among the $982$ approved relationships, 
$88.68\%$ were unanimously labeled as true relationships 
by the contributors and the AROCR was $79.12\%$.
Contributors also contributed to the expansion of the KG 
by providing $838$ new suggestions in total.
Through refinement of these suggestions, 
the maintainers yielded $635$ new and distinct relationships, 
introducing $179$ topics and eight relation types 
that were not previously defined in SED-KGraph.
This made the set of topics grow in size up to $640$.

In the second snapshot, 
the $635$ new relationships acquired 
during the first snapshot 
were subject to evaluation, 
from which $566$ were approved  
and $69$ relationships were deemed ineffective by the contributors.
This yielded a success rate of $89.13\%$ for the Evaluation step.
The AARTR dropped to $74.38\%$ and the AROCR was $72.56\%$.
The reason for such a drop can be explained 
by the number of relation types and their granularity.
The number of relation types reaches $12$, 
which adds to the complexity of the KG 
and might confuse the contributors to some extent.
In this snapshot, the contributors 
made a total of $642$ relationship suggestions.
Upon inspection, the maintainers identified $322$ of these 
as distinct and new ones.
The $322$ new relationships 
introduced $76$ new topics to the KG, 
enlarging the set of topics to $716$ distinct ones.
Moreover, one new relation type was introduced.
Rejected \textit{relation types} (verbs) 
were either too specific (e.g., included versioning) 
or they were synonymous to other verbs. 
The latter, rather than getting rejected, were merged. 
For instance, the two relation types 
\texttt{used-for} and \texttt{provides-functionality} were merged. 
Moreover, rejected \textit{relationships} mainly were rejected 
due to the granularity of the object topic (and in some cases the subject topic). 
For example, if the object topic contained 
a version such as $3$ in \texttt{python3}. 
In such cases, relationships sometimes 
became practically a duplication of another relationship and were rejected or merged.

In the third snapshot, 
we evaluated the new $322$ relationships 
acquired in the previous snapshot 
from which only six were disapproved. 
Contributors verified the remaining $316$ relationships, 
resulting in a success rate of $98.14\%$.
As the KG became more stable and the number of new suggestions dropped, 
the AARTR grows back to an $87.66\%$ and the AROCR is $89.02\%$.
The contributors made a total of $53$ relationship suggestions in this two-months period.
After refinement, the maintainers 
acquired $39$ new and distinct relationship suggestions.
These suggestions introduced $15$ new topics to the KG.

For the final snapshot,
maintainers identified $17$ more topics, 
and acquired $81$ new topics added to GitHub's feature topic list during the past few months.
Maintainers gradually injected these topics, 
without initializing their relationship set, 
into the KG.
We asked contributors to define new relationships for these topics.
As a result of this extra knowledge acquisition step, 
contributors defined $532$ more relationships.
Refinement of these suggested relationships resulted in a set of $372$ relationships, 
introducing $51$ more topics.
These acquired relationships were also gradually injected into the KG 
and evaluated over the same six months.
Therefore, for the fourth snapshot, 
a total of $411$ relationships 
(acquired over the previous step and during the fourth step) were evaluated.
This resulted in $370$ of the relationships under review getting accepted 
and $41$ of them getting rejected, yielding a success rate of $90.02\%$.
We terminated this long-term phase as the number of suggested relationships 
and topics by the contributors gradually diminished.
\begin{table}[tb]
\centering
\caption{Relation types' descriptions and examples}
\begin{tabular}{|c|c|p{50mm}|p{22mm}|}

\hline
\textbf{Step} & \textbf{Relation Type} & \textbf{Description} 

& \textbf{Example} \\ \hline

\textbf{\#1}

& \textit{is-a} &
This is the most basic relation type that allows for the categorization of the topics of the same type together. &
(\texttt{django}, \newline \textit{is-a},\newline \texttt{framework})\\
\cline{2-4}

& \textit{is-used-in-field} &
Relationships of this type map the topic to the field or area it is used in and allow for categorization based on the application field. &
(\texttt{django},\newline \textit{is-used-in-field},\newline \texttt{web-development}) \\
\cline{2-4}

& \textit{provides-functionality} &
Relationships of this type map the topic to the functionality (i.e., the functional purpose of the topic) it provides and allow for categorization based on the functionality of topics. &
(\texttt{django},\newline \textit{provides-functionality},\newline \texttt{backend}) \\
\cline{2-4}

& \textit{works-with} &
Relations of this type map the topic to its dependencies or compatibility constraints. This relation is a bidirectional one, i.e., it matches the topics that work together. &
(\texttt{django},\newline \textit{works-with},\newline \texttt{python})\\
\hline

\multirow{1}*{\textbf{\#2}} &

\textit{is-subset-of} & 
This type of relation allows for hierarchical categorization of topics, 
putting the subject topic under a broader concept (object topic). & 
(\texttt{deep-learning},\newline \textit{is-subset-of},\newline \texttt{neural-network}) \\ 
\cline{2-4}

& \textit{is-based-on} &
Relationships of this type indicate that the creation or development of the subject topic was achieved through use of the object topic. &
(\texttt{archlinux},\newline \textit{is-based-on},\newline \texttt{linux}) \\
\cline{2-4}

& \textit{is-focused-on} &
Relationships of this type emphasize the concepts that the subject topic is concerned with. & 
(\texttt{agile},\newline \textit{is-focused-on},\newline \texttt{flexibility}) \\
\cline{2-4}

& \textit{has-property} &
This type of relation connects the subject topic to meta-data topics. 
The meta-data topics only include well-known and widely used ones. & 
(\texttt{mysql},\newline \textit{has-property},\newline \texttt{open-source}) \\
\cline{2-4}

& \textit{overlaps-with} &
This is a bidirectional relation that links two topics that share some common grounds but are not necessarily interdependent. &
(\texttt{robotics},\newline \textit{overlaps-with},\newline \texttt{ai}) \\
\cline{2-4}

& \textit{provides-product} &
This relation type connects the subject topic as a provider to the products it provides. The provider could be a company, a software system, a tool, or any other entity that creates and provides another entity as a product. &
(\texttt{google},\newline \textit{provides-product},\newline \texttt{flutter}) \\
\cline{2-4}

& \textit{provided-by} &
This is the inverse of the \textit{``provides-product''} relation type 
and keeps the provider and the product connected 
when the provider is the topic of the user's interest. & 
(\texttt{atom},\newline \textit{provided-by},\newline \texttt{github}) \\
\cline{2-4}

& \textit{maintained-by} &
Relationships of this type connect the subject topic to the authorities that maintain the subject topic. &
(\texttt{html},\newline \textit{maintained-by},\newline \texttt{w3c}) \\
\hline

\textbf{\#3} &

\textit{has-license} &
This relation type maps connects the subject topic to its corresponding license. &
(\texttt{backbonejs},\newline \textit{has-license},\newline \texttt{mit-license}) \\
\hline
\end{tabular}

\label{tab:relation_type}
\end{table}

\subsubsection{Resultant KG}
\label{sec:kg_results_characteristics}

\begin{figure}[!tb]
    \centering
    \includegraphics[width=\linewidth,keepaspectratio]{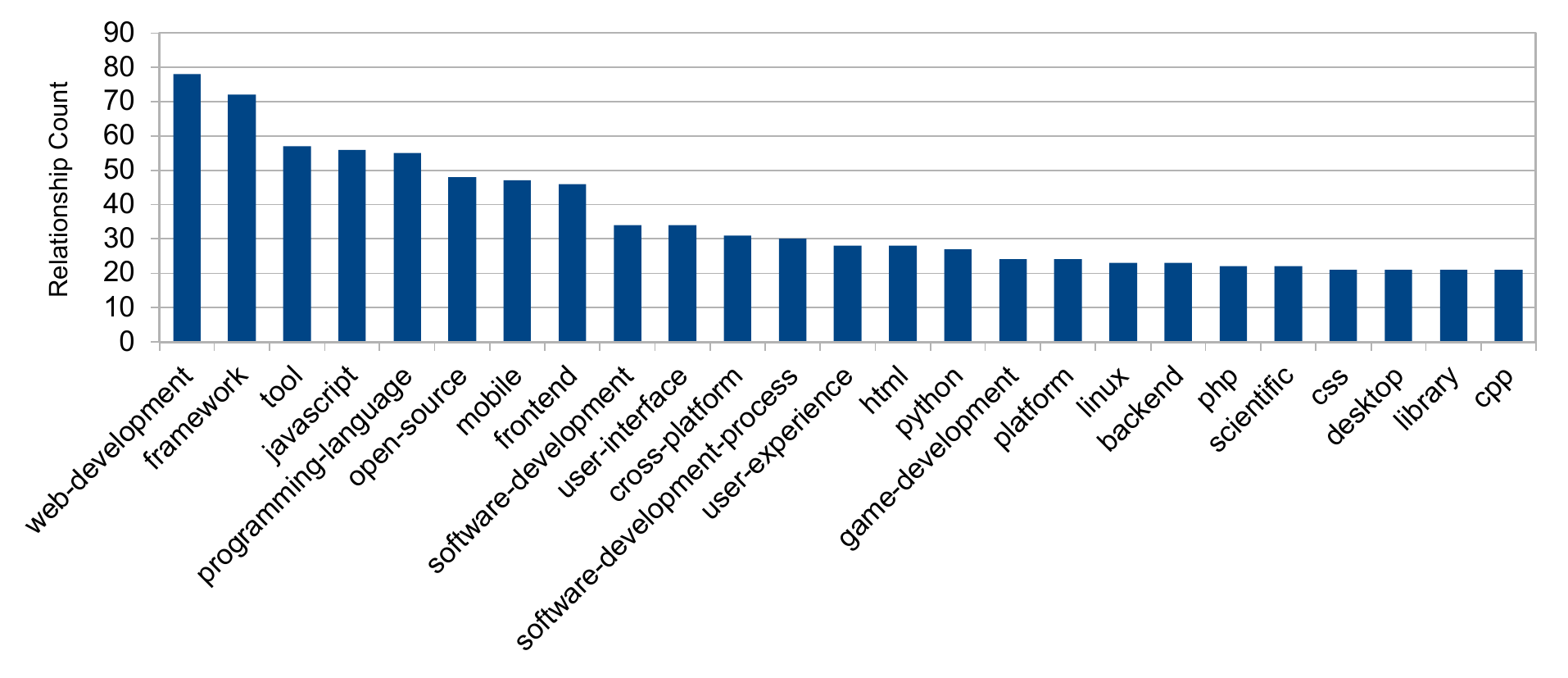}
    \caption{Top 25 most frequent topics}
    \label{fig:topics_tops}
\end{figure}

SED-KGraph consists of \numprint{2234} relationships of 
$13$ relation types among $863$ distinct software topics.
Topics appear as both the subject and the object topic in relationships.
The topic \texttt{web-development} 
has the maximum number of appearances ($78$ relationships), 
while the minimum number of appearances is one.
While one might argue that such rare topics 
should be eliminated from the KG, 
they can be among the useful topics frequently 
used by the community.
Examples of such topics are 
\texttt{awesome}, \texttt{authorization}, and \texttt{augmented-reality}, 
each assigned to \numprint{3863}, \numprint{1847}, 
and \numprint{1628} projects on GitHub, 
making them well-known topics in the community.
They are also evidently important topics in the SED domain.
Not to mention the topics denoting programming languages 
that fall under the same circumstances 
are important topics when used as labels for software entities.
Moreover, we believe that dropping such topics hinders the effective expansion of SED-KGraph.
For SED-KGraph to remain valid and correct, 
it should be continuously expanded as new fields and technologies are always at emergence.
Such rarely present topics might rise to popularity 
or be newly emergent ones that need to be well-established 
in the KG through future contributions.
Thus, we address this issue by assigning weights to topics and relationships.
\autoref{fig:topics_tops} presents the long-tail plot of 
the number of relationship appearances per topic, 
for the $25$ most recurrent ones.
Moreover, \autoref{tab:relation_distrib} details the number of relationships 
in the KG per type of relations.
Notice how the four primary relation types, 
\textit{is-a}, \textit{is-used-in-field}, 
\textit{provides-functionality}, and \textit{works-with}, 
are the most common relationships in the KG.

\begin{table}[!tb]
\centering
\caption{Relation type frequency}
\label{tab:relation_distrib}
\begin{tabular}{lc|lc}
\hline
\textbf{Relation Type} &
\textbf{Count} & 
\textbf{Relation Type} &
\textbf{Count} \\ \hline
\textit{has-license} & 30 & \textit{maintained-by} & 6\\
\textit{has-property} & 134 & \textit{overlaps-with} & 7\\
\textit{is-a} & 578 & \textit{provided-by} & 25\\
\textit{is-based-on} & 55 & \textit{provides-functionality} & 429\\
\textit{is-focused-on} & 43 & \textit{provides-product} & 18\\
\textit{is-subset-of} & 19 & \textit{works-with} & 450\\
\textit{is-used-in-field} & 440 & \\
\hline
\end{tabular}
\end{table}

\subsection{KGRec: Topic Augmentation Model}
Table~\ref{tab:missing_recom_results} 
summarizes the results from the missing topic recommendation task.
As the FCR values indicate, 
TopFilter~\cite{dirocco2020topfilter} 
fails to make any recommendations for almost $50\%$ of the test cases, 
no matter the correctness of the recommendations. 
The reason behind this is mainly the limited number of topics assigned 
to the projects in the dataset ($2.46$ topics on average which is closer to reality).
Any collaborative filtering method 
suffers from the cold start problem~\cite{wang2018ripp}.
An average of $2.26$ topics indicates the data sparsity, 
i.e., there are limited items (topics) assigned to projects, 
which in turn results in the cold start problem.
This shortcoming of TopFilter is also pointed out 
as a limitation of the approach by Di Rocco et al.~\cite{dirocco2020topfilter}.
Taking into account that the dataset is captured 
from a real-world setting, 
this raises questions about the practicality of TopFilter.
However, KGRec overcomes this limitation 
and manages to make recommendations 
under such circumstances.
To better understand the magnitude of the improvements that KGRec brings forth, 
one must take the FCR into account.
The ASR measure for this task is only calculated 
over the set of test cases for which the approach 
under evaluation has managed to make recommendations.
That is, for TopFilter, the ASR is calculated over $54\%$ of the test cases, 
while for KGRec, it is calculated over $98\%$ of the test cases.
Regardless, KGRec outperforms the baselines by $+67\%$ and $+218\%$ 
in terms of ASR and MAP respectively.
That is, KGRec performs considerably better over a wider set of test cases, 
while the baseline has a high ratio of FCR.
\begin{table}[tb]
\centering
\caption{Topic augmentation results}
\label{tab:missing_recom_results}
\begin{tabular}{l|c|ccc}
\hline
\multirow{2}{*}{\textbf{Model}} &
\multicolumn{1}{c|}{\textbf{Over Test Set}} &
\multicolumn{3}{c}{\textbf{Over Sampled Test Set}} \\ \cline{2-5}
& \textbf{FCR} &
\textbf{FCR} &
\textbf{ASR@5} &
\textbf{MAP@5} \\ \hline
Di Rocco et al.\cite{dirocco2020topfilter} & 50.15\% & 46.00\% & 28.33\% & 10.31\% \\ \hline
KGRec (proposed) & \textbf{2.06\%} & \textbf{2.00\%} & \textbf{47.32\%} & \textbf{32.86\%} \\ \hline
\end{tabular}
\end{table}

\subsection{KGRec+: Topic Set Recommendation Model}
To further evaluate KGRec+, 
we improve the baseline approaches 
by combining them  
or stacking KGRec on top of the approach.
We include the resultant models as modified baselines.
Table~\ref{tab:auto_recom_results} 
presents the automated topic set recommendation results.
Notice that reporting FCR for this task is irrelevant 
since the ML-based components of the approaches 
always manage to make a list of recommendations, 
which compose part of the final recommendation lists.
Therefore, FCR for each of the approaches values at zero 
in such circumstances.

The results indicate that
not only does KGRec+ outperforms 
all the previously proposed baselines 
by at least $43.31\%$ and $38.71\%$ in terms of ASR and MAP respectively,
but it also yields better results 
than the modified and improved versions 
of the baseline approaches.
To be exact, KGRec+ outperforms all the approaches, 
including the modified ones 
by at least $25\%$ and $23\%$ in terms of ASR and MAP, respectively.
As mentioned before, 
LR outperforms MNB as the ML-based component.
As we use LR in our approach and to understand 
the impact of KGRec as part of the proposed approach (KGRec+),
one can compare the LR classifier performance 
with and without the KGRec component.
According to this table,  
LR alone achieves $50.8\%$ and $48.93\%$ 
regarding ASR@5 and MAP@5, respectively. 
However, KGRec+ achieves $72.8\%$ and $67.8\%$ 
for ASR@5 and MAP@5, respectively. 
This outperformance ($43\%$ ASR@5 and $39\%$ MAP@5) 
highlights the contribution of KGRec as part of KGRec+.
We believe our unique advantage lies in the fact that current recommenders are restricted to the training data 
which suffers from missing topics. 
Machine learning techniques' performance is usually limited by the quality of the data they consume. 
However, using the knowledge graph built on human expertise, 
we provide the recommender model with missing blocks of information 
that it can exploit to complement and enhance the recommendation list. 
In other words, 
the combination helps us utilize 
both the strengths of the ML-based component 
and expert knowledge.
\begin{table}[!tb]
\centering
\caption{Automated topic set recommendation results}
\begin{tabular}{|c|c|c|c|}
\hline

\multicolumn{2}{|c|}{\textbf{Model}} &
\textbf{ASR@5} &
\textbf{MAP@5} \\ 
\hline

\multirow{2}*{Baselines}
& MNB~\cite{di2020multinomial} & 30.80\% & 27.07\% \\
& TopFilter~\cite{dirocco2020topfilter} & 41.20\% & 33.47\% \\
& LR~\cite{izadi2021topic} & 50.80\% & 48.93\% \\
\hline

\multirow{2}*{\pbox{10mm}{Modified\\Baselines}}
& KGRec plus MNB~\cite{di2020multinomial} & 54.80\% & 42.94\% \\
& TopFilter~\cite{dirocco2020topfilter} plus LR~\cite{izadi2021topic} 
& 58.40\% & 55.05\% \\
\hline

Proposed
& KGRec+ (KGRec plus LR~\cite{izadi2021topic})  
& \textbf{72.80\%} & \textbf{67.87\%}\\ \hline

\multicolumn{2}{|l|}{\textbf{\pbox{35mm}{Outperforming baselines by}}}
& \textbf{+25\% to +136\%} & \textbf{+23\% to +151\%} \\
\hline

\label{tab:auto_recom_results}
\end{tabular}
\end{table}

\subsection{Discussion}
In the following, 
we discuss 
various aspects of our approach and its settings, 
our results, 
lessons learned, 
and possible applications of this work.

\paragraph{KGRec+ Parameter Setting}
When recommending the final top $k$ topics per repository, 
we take $m$ topics from the ML-based component 
and $g$ topics from the KGRec component, 
where $k = m + g$. 
We can use arbitrary values of $m$ and $g$ that fit the above criterion.
Increasing $m$ translates to taking more topics from the ML-based component. 
But one should also consider 
the number of available topics in the ground truth set 
on which the model is trained 
to avoid going beyond the characteristics/limitations of the classifier. 
Moreover, in a fixed-length recommendation list, 
increasing $m$ leaves fewer places 
for missing topics to be predicted, 
hence hindering effective evaluation of 
the missing topic recommender component. 
On the other hand, decreasing $m$ 
causes the approach to rely on fewer topics 
discovered by the ML-based component. 
Note that providing very few initial topics 
may cause the missing topic recommender component 
to struggle for finding sufficient relevant topics. 
Hence, all this should be taken into account 
while tuning these parameters' values in different use cases. 
For our application, 
as the average number of topics per repository 
is $2.4$, and the model is trained on that data, 
we choose $m = 3$ to be fair to the ML-based component. 
Then, as we aim to complement the predicted topic set, 
we set $g = 2$ to construct a set of $5$ topics per repository. 
This setting considering the dataset characteristics seemed more logical. 
Regarding the total size of the recommended list, 
we also experimented with other settings. 
As expected, as the number of predicted tags goes up, 
the recall score increases and precision decreases. 
This means while we are becoming more confident 
that the ground truth tags are being retrieved by the recommender, 
more unrelated or missed tags 
can be also added to the recommendation list. 
This highlights the need for the manual inspection of 
these missing topics to determine whether 
they are relevant to the repository or not. 
Hence, this number needs to be customized 
based on the dataset and the problem domain 
at hand in other use cases.

\paragraph{Tangled Topics}
In this study, 
we mainly focused on constructing the KG 
using the help of experts 
to avoid tangled topics and tag explosion. 
Hence, we set a number of policies in our platform 
to prevent tangled topics 
polluting the data as much as possible. 
More specifically, in our platform,
we have redundancy detection, spam report, and edit suggestion capabilities 
to mitigate such risks.
Please refer to Section~\ref{sec:platform} 
for the platform functionalities. 
Redundancy detection can potentially detect tangled topics. 
When someone defines a topic, 
they can see the list of the redundancies 
with that new topic 
and either delete their topic 
to resolve the redundancy 
or edit it to clarify the differences if needed. 
Maintainers can also check reported redundancies 
and resolve them. 
The same goes for the tag explosion problem. 
Finally, we would like to point out
that there are more advanced features 
to detect semantically-similar topics 
such as ML techniques using contextual embeddings. 
This is indeed a possible future feature for our platform.

\paragraph{Lessons Learned}
Through our experiments, we learned a few lessons;
some helped us better understand and adjust our process, 
and some are worthy of further investigation by future studies.
Next, we will review them.
Based on our experience in this study, 
we realized that topic sets 
can easily enlarge and become irrelevant or useless 
for practical downstream tasks. 
Hence, it was evident to us that 
to capture a high-quality set of topics, manual inspection by experts is highly recommended. 
However, one should also mind the cost. 
Here, semi-automatic features can be helpful.
Another clear lesson was the fact that 
the KG should be maintained and updated 
as the SE community grows and expands.
Hence, we improved our platform 
to include several automatic and semi-automatic functionalities 
to help maintain the KG.
Moreover, we observed that SE practitioners
seem to be more familiar with SE topics 
and how they work or relate to each other compared to SE researchers
without any industrial background. 
This may be due to the fact that practitioners 
interact with these technologies on a daily basis 
and \textit{may} be more up-to-date. 
This notion can indicate that topics in our KG 
may be more practical in nature. 
This is of course not verified and 
a controlled experiment and more qualitative studies 
are required to confirm such assumptions. 
Based on the assessments, 
we also suggest that contributors 
limit their contributions to topics related to their specialty 
rather than trying to contribute to the whole graph. 
We emphasize this notion to our contributors on the platform. 
An interesting remark is that
some concepts or fields may be lacking initial seed topics to begin with. 
For instance, a popular concept 
such as software security does not have well-curated seed topics 
in GitHub's featured set yet. 
This makes it more difficult for contributors 
to expand the KG for this specific field 
as they rarely come across topics hinting at security. 
Subsequently, recommenders based on this may under-recommend such topics. 
This is an existing challenge for such applications and is an interesting line of research to pursue. 
Finally, we observed that as the topic set matures and becomes more stable over time, 
the growth in the average performance of recommenders over all topics slows down gradually. 
This can indicate that while the KG helps build better recommenders 
(compared to recommenders that do not utilize the KG), 
individual repositories associated with newer or rarely-used topics 
may benefit more from the growth of the KG over time.

During the KG construction and expansion phases, 
we encountered a few controversial aspects of topics/relationships including:
A main controversial issue was 
to whether use abbreviated or long forms of topics. For instance, we can use \texttt{SE} instead of \texttt{Software Engineering}. 
Currently, our policy is to include both 
under the display and full names for a topic to avoid confusion or redundancy. 
Moreover, some topics have different meanings. 
This is exasperated in the case of abbreviated topics. 
In these cases, one can refer 
to tagged repositories to determine the difference. 
Users who try to introduce such topics 
will be notified of duplicates. 
Moreover, topics that are a single topic in nature may seem to be compound at first look. For example, \texttt{material-design-for-bootstrap} 
sounds compound but it is a single topic. 
Another controversy may arise 
from the experience or familiarity level of users 
with different SE fields.
Consider the topic \texttt{less} which stands for Leaner Style Sheets
in the SE domain and refers to a dynamic preprocessor style sheet language 
that can be compiled into CSS. 
However, an inexperienced user may confuse it with the ``less" determiner in the English language. 
Finally, bidirectional relationships such as \texttt{works-with} 
may create duplicate pairs. 
Nonetheless, the multiple-rater policy 
we have in place on the platform helps avoid mistakes.

\paragraph{Implications and Potential Applications}
We curated SED-KGraph to help 
build better topic recommenders for software repositories. 
Through experiments, 
we demonstrated that utilizing SED-KGraph 
in the topic recommendation task 
improves the quality of recommendation lists. 
Topic sets assigned to repositories 
can be enlarged twice their original size 
using only cleaned and high-quality topics. 
Moreover, the accuracy of complemented topic sets 
can be increased up to 151\% which is a notable improvement. 
Moreover, it has been shown that the correctness and completeness 
of the set of topics assigned to entities improve their visibility, 
and in turn, impact the performance of any topic-based solution 
to information retrieval problems~\cite{held2012learning}. 
Hence, SED-KGraph can also be beneficial in other settings. 
Our work paves the way 
for future research with many possible directions 
to improve both industrial and academic problems. 
Practitioners and researchers can tailor our KG 
to other applications including (but not limited to) 
categorizing SE entities, 
enhancing retrieval and searching, 
improving exploratory navigation in information websites, 
recommending similar repositories, 
discovering duplicate QA posts on Stack Overflow, and many more. 
Furthermore, SED-KGraph, as a structured knowledge base, 
can serve as a rich source of information on SED topics themselves. 
In SED-KGraph, topics are stored in the form of info boxes, 
i.e., topics are saved along with a list of their aliases, 
links to the informative pages on the topics, 
and short descriptions of the topics.
Therefore, information on SED topics 
is easily accessible through queries, 
especially since the semantics of relation types 
can be easily injected into the queries. 
We hope the SE community utilizes this carefully 
curated knowledge base for numerous downstream tasks.


\subsection{Threats to Validity}\label{sec:threats}
In this section, we discuss the possible threats 
to the validity of this study and how we have addressed these threats.

\paragraph{Internal Validity}
\label{sec:threats_internal}
These threats correspond to the correctness of the relationships
and the subjectiveness of contributors and maintainers.
We address the former by evaluating every relationship 
through a partially objective labeling task 
in which at least three and up to nine contributors 
validate the correctness of the relationships.
As for the latter, aside from the KG Initialization stage, 
the effect of maintainers' personal experience and knowledge 
was minimized by limiting their role to fixing consistency issues in the suggestions 
from the community in the continuous second stage.
The subjectiveness of the contributors was also mitigated
by engaging $170$ experts as contributors.
Although we invited people with relevant and sufficient background 
in SE and development,
it is possible that some participants may not be familiar with some topics.
To avoid inaccurate votes, 
we make it possible for contributors 
to only contribute to the topics related to their expertise 
and skip unfamiliar topics.
Moreover, we provided background information on topics 
so that the participants can make more informed contributions.
Another factor can be errors in our source code.
We have double-checked the source code to decrease this threat,. 
However, there could be experimental errors in the setup that we did not notice.
Therefore, we have publicly released our source code and dataset, 
to enable the community to use and/or replicate our work\footnote{\url{https://github.com/mahtab-nejati/KGRec}}.

\paragraph{External Validity}\label{sec:threats_external}
These threats correspond to the generalizability and effectiveness 
of our graph and recommenders. 
Through crowd-sourcing and the expansion of SED-KGraph, 
we address the generalizability and effectiveness concerns.
We also validated the relationships in multiple snapshots with the community, 
assuring their correctness and effectiveness.
Although we use GitHub's featured set as the initial seed set, 
the resultant KG is not restricted to any platform 
and can be reused in other software-related platforms.
Contributors were indeed instructed 
to incorporate their knowledge of Software Engineering 
while assessing/suggesting relationships and topics, 
irrespective of any specific SED platform.
As for the topic recommendation, 
the KGRec component only takes the initial set of topics 
assigned to software projects as the input.
Thus, it can be easily adapted for use on any software-related platform and any software entity.
For the KGRec+ model, 
as long as proper textual information is available for a project, 
the model is able to recommend relevant topics.
Moreover, the ML-based component can be re-trained 
with textual information of other software entities, 
using the model to recommend
topics for those entities as well.
Also for training, 
datasets were randomly split to avoid introducing bias.
Finally, when assessing recommenders' performance,
we use $50$ repositories.
Incorporating more repositories 
can improve the generalizability of our approach.
However, assessing the correctness of the assigned missing tags 
requires \textit{manual} inspection per repository and approach. 
Not to mention that each sample is examined by multiple evaluators 
to avoid introducing bias. 
This accumulates to a large number of evaluations and takes a lengthy time. 
Due to this, we take at most $50$ repositories 
and could not afford to increase the number of repository samples.

\paragraph{Construct Validity}\label{sec:threats_construct}
These threats correspond to the features and capabilities of the online platform 
used for KG expansion, and the sensitivity of KGRec to its input.
We allowed for free-text topics and relation types in the suggestion forms to allow for the expansion of the KG.
This resulted in consistency issues, 
which maintainers handled by refining the suggestions.
This concern is further handled in the final version of the platform 
through the use of specially designed features and policies.
Moreover, KGRec is sensitive 
to the correctness of the initial set of topics 
assigned to the project, 
such that the inaccuracy misleads the model on SED-KGraph.
This limitation calls for an accurate ML-based component 
to be combined with KGRec.
To address this threat, 
we used a multi-class multi-label LR classifier, 
which has been shown to exhibit the best performance 
among similar ML-based approaches~\cite{izadi2021topic}.

\section{Related Work}
\label{sec:related}
We organize the related work as approaches on 
(1) topic recommendation for software projects, 
(2) for other software entities, and finally studies on
(3) KGs for software engineering.

\paragraph{Topic Recommendation for Software Projects:}
There are several studies with a focus on 
the topic recommendation for software projects~\cite{xia2013tag,xia2015tagcombint,vargas2015automated,cai2016greta,zhou2017scalable,wang2018entagrec++,wang2014tag,liu2018fasttagrec,di2020multinomial,izadi2021topic}.
\textit{Sally} presented by Vargas-Baldrich et al.~\cite{vargas2015automated}, is a tool for generating topics for Maven-based software projects by analyzing their bytecode and the dependency relations among them.
Their approach, unlike ours, is limited in application due to being dependent on the programming language.
Cai et al.~\cite{cai2016greta} proposed 
a graph-based cross-community approach called\textit{GRETA}, 
for assigning topics to repositories.
Their approach is to first construct an Entity-Tag Graph 
and for each queried project, 
take a random walk on a subset of the graph around the most similar entities 
to the queried one to assign tags to the project.
While they do propose a graph-based approach, 
their graph fundamentally differs from ours in nature.
Di Sipio et al.~\cite{di2020multinomial},
proposed using an MNB classifier for the classification of about $134$ GitHub topics. 
In each top-$k$ recommendation list, 
authors predict $k-1$ topics using text analysis and one topic 
using a code analysis tool called \textit{GuessLang}.
TopFilter~\cite{dirocco2020topfilter},
the state-of-the-art for missing topic recommendation,
is the most similar study to ours in terms of purpose (topic augmentation task).
Authors take an item-based collaborative approach for recommending missing topics.
Our experiments prove that their approach suffers from practicality issues, 
which our proposed approach overcomes.
Most recently, Izadi et al.~\cite{izadi2021topic},
demonstrated the impact of clean topics and
proposed a multi-label Logistic Regression classifier for recommending topics.
Our approach is orthogonal to this study, 
thus we incorporate their approach in this work.

\paragraph{Topic Recommendation for other Software Entities:}
There are several pieces of research on tag recommendation 
for other types of software entities such as questions on Stack Overflow, Ask Ubuntu, and Ask Different~\cite{wang2018entagrec++,wang2014tag,zhou2017scalable,xia2013tag,liu2018fasttagrec,maity2019deeptagrec}. 
The discussion around these tags and their usability in the SE community 
have been so fortified that 
the Stack Overflow platform has also developed a tag recommendation system of its own.
These approaches mostly employ word similarity-based and semantic similarity-based techniques.
Xia et al.~\cite{xia2013tag} focused on 
calculating the similarity based on the textual description. 
The authors propose \textit{TagCombine} 
for predicting tags for questions using 
a multi-label ranking method based on OneVsRest Naive Bayes classifiers. 
Semantic similarity-based techniques \cite{wang2018entagrec++,wang2014tag,liu2018fasttagrec} 
consider text semantic information and perform significantly better than the former approach.
Wang et al.~\cite{wang2018entagrec++,wang2014tag}, 
proposed \textit{ENTAGREC} and \textit{ENTAGREC++} 
which uses a mixture model based on LLDA to consider all tags together. 
Liu et al. \cite{liu2018fasttagrec}, proposed \textit{FastTagRec}, 
for tag recommendation using 
a neural-network-based classification algorithm and bags of n-grams (bag-of-words with word order).
As a possible future direction, 
our SE-based KG and approach can also be utilized 
for recommending topics for these types of software entities.

\paragraph{KGs for Software Engineering:}

KGs have been utilized in numerous studies 
to address different software engineering problems.
In some cases, researchers strive for a fully automated KG extraction approach 
from the available textual data.
However, more often than not, 
these studies narrow the scope of their KG's content 
down to very specific aspects.
In such cases, concept extraction from the textual data can be achieved 
through use case-specific tailored solutions, 
especially when the input data for KG construction is in a semi-structured or expected format~\cite{li2018api,chen2019bug,sun2019knowhow,sun2020taskoriented,wang2017construct} 
or in some extreme cases, the concepts are predefined\cite{zhao2019gitrepo}.
Some studies recognize that a fully-automated approach 
does not suffice their purpose, 
even in the limited scope of knowledge they intend to model 
and opt for semi-automated solutions~\cite{karthik2019latent,cao2019coding}.
HDSKG~\cite{zhao2017hdskg} 
is a framework for mostly-automated KG construction.
The authors applied their approach 
to the \textit{tagWiki} pages on Stack Overflow 
in an attempt to construct a domain-specific KG of SED topics.
The authors claim HDSKG includes 
\numprint{44800} unique concepts (topics) 
and \numprint{35279} relation triples of 
\numprint{9660} unique verb phrases (relation types).
While HDSKG can guarantee 
the lexical uniqueness of concepts and verb phrases
through applying text processing techniques,
the semantic uniqueness can not be promised.
The lack of semantic uniqueness leads to
tag explosion and redundant/duplicate verb phrases and relationships, 
which can be well hidden since neither the concepts nor the verb phrases 
are mapped to their semantically equivalent terms.
Unfortunately, neither the resultant KG 
nor the code base of this work is publicly available.
However, based on the sample nodes of HDSKG, 
its automatic method of extracting noun phrases results in tangled topics such as \texttt{small-java-library}.
This justifies the enormous number of extracted topics and relationships.
While HDSKG can be used in conjunction with our approach 
and replace the Acquisition process, 
the sheer multitude of the concepts and verb phrases works 
to the detriment of integrity and consistency concerns.
The applicability of a KG of SED topics 
is highly sensitive to tag explosion and tangle topics problems, 
a threat that semi-automated and fully-automated approaches fail to mitigate.
Especially with tangled topics as a concern 
while there are compound topics that convey atomic concepts 
as SED topics, each of the approaches leans towards detecting one and missing the other.
Finally, some studies resort to fully-manual construction approaches 
due to data scatteredness and sparsity~\cite{fathalla2018eventskg} 
or make use of pre-constructed community-defined KGs 
for their purposes~\cite{han2018deepweak}.
Consequently, we opted for a hybrid approach to avoid the pitfalls of each method described above, 
while obtaining high-quality topics and relationships as much as possible.

\section{Future Work}
\label{sec:future}

The main contributions of this study 
namely the SED-KGraph, 
the KGRec+ topic recommender, 
and also the platform 
all can be improved with future work. 
In the following, 
we present possible directions 
for extending this work.
In the future, 
we aim to maintain both SED-KGraph and the platform through 
which is expanded by the community.
We can further improve different aspects of our platform 
and add more automated solutions.
For instance, we aim to 
enhance our redundancy checker using ML techniques 
to discover semantically-similar topics that are written differently.
The KG itself and the contributions received from the community
can be further investigated.
For instance, several factors 
including participants' numbers, 
duration of the snapshot, 
gender or background of participants, etc. 
differ for different snapshots.
Such factors may impact the results of various snapshots. 
Further studies are required to explore the possible effects.
The ML-based component for recommending topics 
can also be improved. 
For instance, one can also process the source files 
associated to repositories to enhance the ability of the classifiers 
and suggest more relevant topics.
Moreover, one can invest in training contextual-based models 
to further improve the performance of the recommenders.
Currently, our missing topic recommender does not utilize the relation types. 
Different relation types and their frequencies, 
if properly investigated, 
can potentially help build stronger recommenders.
This is also a possible direction to investigate in future research.
Finally, we aim to study other applications of SED-KGraph 
in different contexts such 
as search engines and information retrieval 
as well as other information websites such as Stack Overflow.

\section{Conclusions}
\label{sec:conclude}
To discover the semantic relationships 
among SED-related topics,
we engaged $170$ researchers and practitioners 
to collaboratively construct 
a  KG we call SED-KGraph.
To initialize the KG, 
we first drafted a primary version, 
taking GitHub's featured topics 
as the seed topic set.
Through their contributions, we constructed SED-KGraph 
with \numprint{2234} carefully evaluated relationships 
among $863$ community-curated topics.
We also developed a platform through which 
we evaluated and expanded SED-KGraph in a crowd-sourced continuous method.
Knowing that the KG will keep growing as do the SED technologies, 
we maintained the platform to sustain the continuous expansion of SED-KGraph 
in a more continuous and semi-automated manner requiring less human effort.

In the second stage of the proposed approach, 
we propose two recommender systems, 
KGRec and KGRec+ for tagging software projects
augmented by the semantic relationship among their topics.
We developed KGRec to predict the missing topics 
of software projects in GitHub based on SED-KGraph.
The second recommender, however, assumes there is no topic available 
for a repository, and proceeds to predict the relevant topics 
based on both textual information of a software project 
(such as its README file), SED-KGraph, 
and its ML-based component.
Our experiments yield that this model achieves $1.7X$ and $3.2X$ 
higher scores regarding ASR$@5$ and MAP$@5$, respectively.
We also built upon the missing topic recommender (KGRec) 
and added an ML-based component to the approach 
to develop a stand-alone automated topic recommender system, KGRec+.
The results show that KGRec+ 
outperforms the state-of-the-art baseline approaches 
as well as the modified and improved ones by at least $+25\%$ and $+23\%$ 
regarding ASR$@5$ and MAP$@5$ measures, respectively.
Finally, we publicly share SED-KGraph, 
as a rich form of knowledge for the community to reuse and build upon.
Furthermore, we release the source code of our two recommender models.

\begin{acknowledgements}
We would like to thank all the participants 
for helping us with constructing and evaluating our KG, 
as well as for assessing our recommender model.
\end{acknowledgements}

\section*{Conflict of interest}
The authors declare that they have no conflict of interest.

\section*{Data Availability Statements}
The dataset generated during the current study is available in the authors’ public GitHub repository.~\footnote{\url{https://github.com/mahtab-nejati/KGRec}} 
The dataset used for training the ML-based components and comparing approaches is also available in the baseline paper’s public GitHub repository.~\footnote{\url{https://github.com/MalihehIzadi/SoftwareTagRecommender}}

\begin{appendices}
\section{Platform Screenshots}
\label{appendix:platform}

\begin{figure}
    \centering
    \begin{subfigure}[b]{0.48\textwidth}
    \centering
        \includegraphics[width=\linewidth]{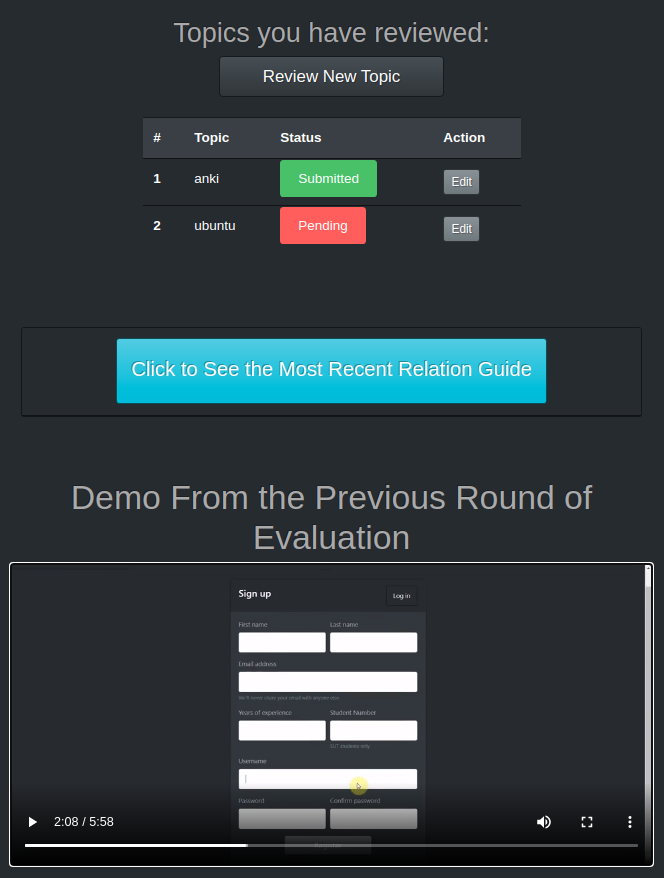}
        \caption{Dashboard}
        \label{fig:dashboard}
    \end{subfigure}
    \begin{subfigure}[b]{0.48\textwidth}
    \centering
        \includegraphics[width=\linewidth]{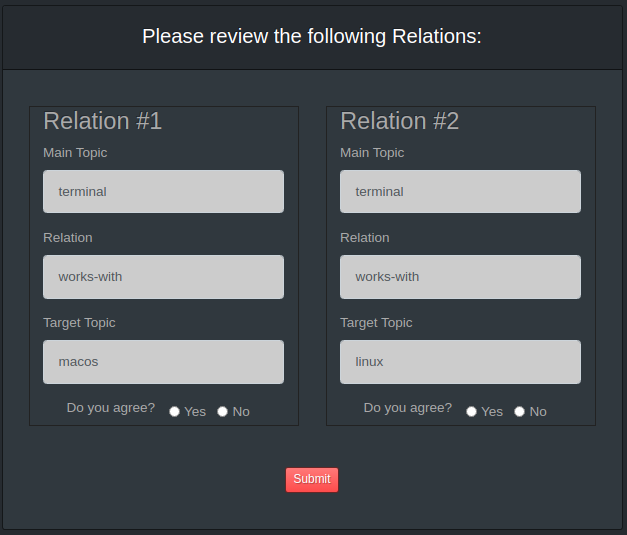}
        \caption{Review a relationship}
        \label{fig:review_form}
    \end{subfigure}
    \caption{Platform dashboard and review panel for KG construction phase}
    \label{fig:platform_phase1}
\end{figure}

\begin{figure}
    \centering

    \begin{subfigure}[b]{0.49\textwidth}
        \centering
        \includegraphics[width=\linewidth]{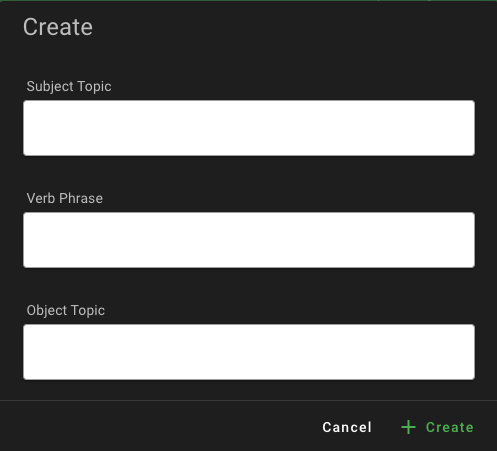}
        \caption{Define new relationship}
        \label{fig:define_relationship}
    \end{subfigure}
    \begin{subfigure}[b]{0.49\textwidth}
        \centering
        \includegraphics[width=\linewidth]{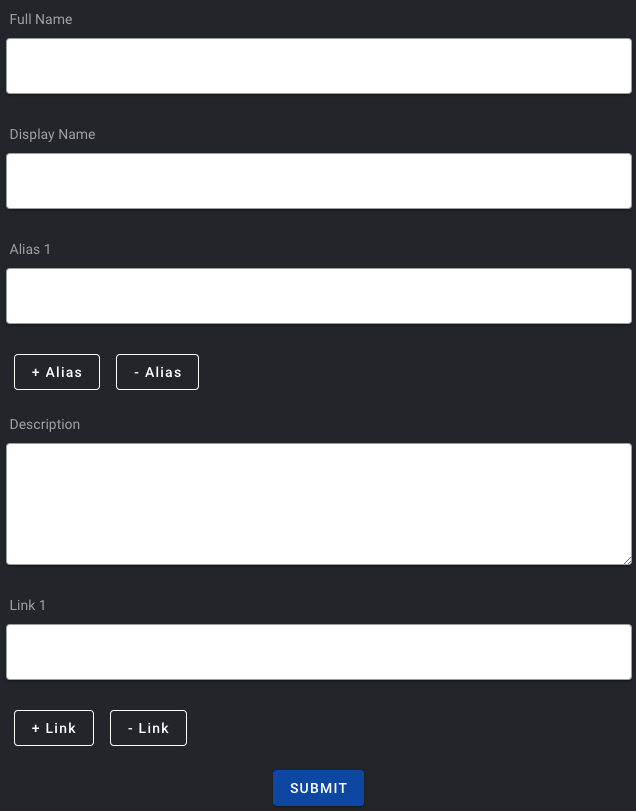}
        \caption{Define new topic}
        \label{fig:define_topic}
    \end{subfigure}
    \begin{subfigure}[b]{0.49\textwidth}
        \centering
        \includegraphics[width=\linewidth]{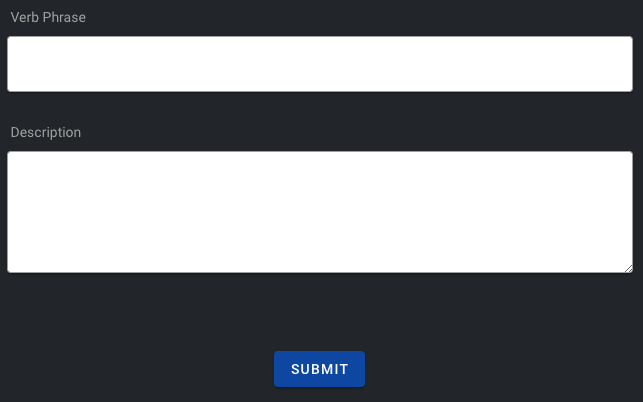}
        \caption{Define new relation type}
        \label{fig:define_verb}
    \end{subfigure}
        \begin{subfigure}[b]{0.49\textwidth}
        \centering
        \includegraphics[width=\linewidth]{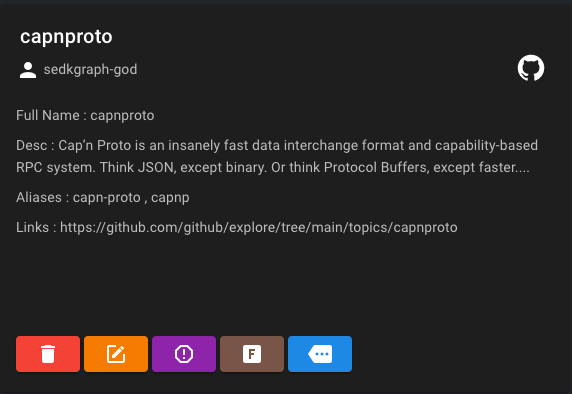}
        \caption{A sample topic}
        \label{fig:sample_topic}
    \end{subfigure}
    \begin{subfigure}[b]{0.49\textwidth}
        \centering
        \includegraphics[width=\linewidth]{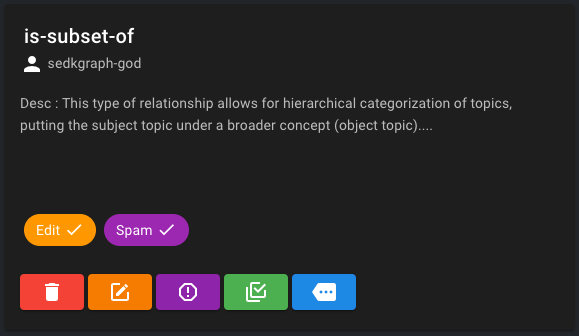}
        \caption{A sample relation type}
        \label{fig:sample_verb}
    \end{subfigure}
        \begin{subfigure}[b]{0.49\textwidth}
        \centering
        \includegraphics[width=\linewidth]{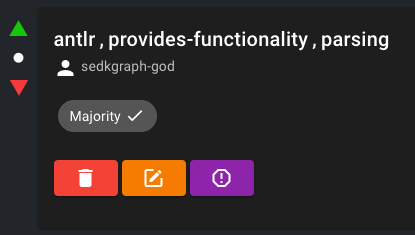}
        \caption{A sample relationship}
        \label{fig:sample_relationship}
    \end{subfigure}
    \caption{KG entities in the maintenance phase}
    \label{fig:platform_creations}
\end{figure}

Figures~\ref{fig:platform_phase1} and 
\ref{fig:platform_creations} 
present multiple screenshots 
of our online platform 
for both the construction and maintenance phases.


\section{Samples}
\label{appendix:sample_rels}
Table~\ref{tab:sample_relation_type}
presents several samples per relation type from SED-KGraph.

\begin{table}[tb]
\centering
\caption{Samples per relation types}
\begin{tabular}{|c|p{70mm}|}

\toprule
\textbf{Relation Type} & \textbf{Samples} \\ 
\midrule

\textit{is-a} &
(\texttt{django}, \textit{is-a}, \texttt{framework}) \newline
(\texttt{android}, \textit{is-a}, \texttt{operating-system}) \newline
(\texttt{atom}, \textit{is-a}, \texttt{text-editor})
\\\midrule

\textit{is-used-in-field} &
(\texttt{django}, \textit{is-used-in-field}, \texttt{web-development}) \newline
(\texttt{3d}, \textit{is-used-in-field}, \texttt{graphics}) \newline
(\texttt{azure}, \textit{is-used-in-field}, \texttt{cloud-computing}) 
\\\midrule

\textit{provides-functionality} &
(\texttt{django}, \textit{provides-functionality}, \texttt{backend})\newline
(\texttt{auth0}, \textit{provides-functionality}, \texttt{authentication})\newline
(\texttt{blockchain}, \textit{provides-functionality}, \texttt{decentralization})
\\\midrule

\textit{works-with} &
(\texttt{django}, \textit{works-with}, \texttt{python})\newline
(\texttt{blockchain}, \textit{works-with}, \texttt{cryptography})\newline
(\texttt{kubernetes}, \textit{works-with}, \texttt{docker})
\\\midrule

\textit{is-subset-of} & 
(\texttt{image-processing}, \textit{is-subset-of}, \texttt{machine-learning})\newline 
(\texttt{continuous-deployment}, \textit{is-subset-of}, \texttt{cicd})\newline 
(\texttt{user-experience}, \textit{is-subset-of}, \texttt{ui-ux})
\\ \midrule

\textit{is-based-on} &
(\texttt{archlinux}, \textit{is-based-on},\texttt{linux}) \newline 
(\texttt{xmake}, \textit{is-based-on},\texttt{lua}) \newline 
(\texttt{reactiveui}, \textit{is-based-on},\texttt{mvvm}) 
\\\midrule

\textit{is-focused-on} & 
(\texttt{agile}, \textit{is-focused-on}, \texttt{speed})  \newline
(\texttt{end-to-end-encryption}, \textit{is-focused-on}, \texttt{privacy})  \newline
(\texttt{neo4j}, \textit{is-focused-on}, \texttt{scalability})  
\\\midrule

\textit{has-property} & 
(\texttt{mysql}, \textit{has-property}, \texttt{open-source}) \newline
(\texttt{anki}, \textit{has-property}, \texttt{cross-platform}) \newline
(\texttt{elite-dangerous}, \textit{has-property}, \texttt{multiplayer})
\\\midrule

\textit{overlaps-with} &
(\texttt{robotics}, \textit{overlaps-with}, \texttt{ai}) \newline
(\texttt{data-science}, \textit{overlaps-with}, \texttt{ai}) \newline
(\texttt{nlp}, \textit{overlaps-with}, \texttt{machine-learning}) 
\\\midrule

\textit{provides-product} &
(\texttt{google}, \textit{provides-product}, \texttt{flutter})\newline
(\texttt{amazon}, \textit{provides-product}, \texttt{aws})\newline
(\texttt{mediawiki}, \textit{provides-product}, \texttt{wikipedia})
\\\midrule

\textit{provided-by} & 
(\texttt{atom}, \textit{provided-by}, \texttt{github}) \newline
(\texttt{flutter}, \textit{provided-by}, \texttt{google}) \newline
(\texttt{macos}, \textit{provided-by}, \texttt{apple}) 
\\\midrule

\textit{maintained-by} &
(\texttt{html}, \textit{maintained-by}, \texttt{w3c}) \newline
(\texttt{symfony}, \textit{maintained-by}, \texttt{sensiolabs-sas})  \newline
(\texttt{uportal}, \textit{maintained-by}, \texttt{apereo}) 
\\\midrule

\textit{has-license} &
(\texttt{backbonejs}, \textit{has-license}, \texttt{mit-license}) \newline
(\texttt{ansible}, \textit{has-license}, \texttt{gnu-gpl-license}) \newline
(\texttt{robotframework}, \textit{has-license}, \texttt{apache-license}) 
\\
\bottomrule
\end{tabular}

\label{tab:sample_relation_type}
\end{table}

\end{appendices}

\bibliographystyle{abbrv} 
\bibliography{main}

\end{document}